\documentclass[graybox]{svmult}

\usepackage{mathptmx}       
\usepackage{helvet}         
\usepackage{courier}        
\usepackage{type1cm}        
\usepackage{makeidx}         
\usepackage{graphicx}        
\usepackage{multicol}        
\usepackage[bottom]{footmisc}

\makeindex             

\begin{document}

  \newcommand {\nc} {\newcommand}
  \nc {\beq} {\begin{eqnarray}}
  \nc {\eeq} {\nonumber \end{eqnarray}}
  \nc {\eeqn}[1] {\label {#1} \end{eqnarray}}
  \nc {\eol} {\nonumber \\}
  \nc {\eoln}[1] {\label {#1} \\}
  \nc {\mrm} [1] {\mathrm{#1}}
  \nc {\half} {\mbox{$\frac{1}{2}$}}
  \nc {\thal} {\mbox{$\frac{3}{2}$}}
  \nc {\fial} {\mbox{$\frac{5}{2}$}}
  \nc {\la} {\mbox{$\langle$}}
  \nc {\ra} {\mbox{$\rangle$}}
  \nc {\etal} {\emph{et al.\ }}
  \nc {\eq} [1] {(\ref{#1})}
  \nc {\Eq} [1] {Eq.~(\ref{#1})}
  \nc {\Ref} [1] {Ref.~\cite{#1}}
  \nc {\Refc} [2] {Refs.~\cite[#1]{#2}}
  \nc {\Sec} [1] {Sec.~\ref{#1}}
  \nc {\chap} [1] {Chapter~\ref{#1}}
  \nc {\anx} [1] {Appendix~\ref{#1}}
  \nc {\tbl} [1] {Table~\ref{#1}}
  \nc {\fig} [1] {Fig.~\ref{#1}}
  \nc {\ex} [1] {$^{#1}$}
  \nc {\Sch} {Schr\"odinger }
  \nc {\dem} {\mbox{$\frac{1}{2}$}}
  \nc {\flim} [2] {\mathop{\longrightarrow}\limits_{{#1}\rightarrow{#2}}}
  \nc {\textdegr}{$^{\circ}$}

\title*{Breakup reaction models for two- and three-cluster projectiles}
\author{D. Baye and P. Capel}
\institute{D. Baye \at Physique Quantique, C.P. 165/82 and 
Physique Nucl\'eaire Th\'eorique et Physique Math\'ematique, C.P. 229,
Universit\'e Libre de Bruxelles, B 1050 Brussels, Belgium, \email{dbaye@ulb.ac.be}
\and P. Capel \at National Superconducting Cyclotron Laboratory,
Michigan State University, East Lansing MI-48824, USA,
\email{capel@nscl.msu.edu}}
\maketitle

\abstract*{Breakup reactions are one of the main tools 
for the study of exotic nuclei, and in particular of their continuum. 
In order to get valuable information from measurements, 
a precise reaction model coupled to a fair description 
of the projectile is needed. 
We assume that the projectile initially possesses a cluster structure, 
which is revealed by the dissociation process. 
This structure is described by a few-body Hamiltonian involving effective 
forces between the clusters. 
Within this assumption, we review various reaction models. 
In semiclassical models, the projectile-target relative motion is described 
by a classical trajectory and the reaction properties are deduced by solving 
a time-dependent \Sch equation. 
We then describe the principle and variants of the eikonal approximation: 
the dynamical eikonal approximation, 
the standard eikonal approximation, 
and a corrected version avoiding Coulomb divergence. 
Finally, we present the continuum-discretized coupled-channel method (CDCC), 
in which the \Sch equation is solved with 
the projectile continuum approximated by square-integrable states. 
These models are first illustrated by applications to two-cluster projectiles 
for studies of nuclei far from stability and of reactions useful in astrophysics. 
Recent extensions to three-cluster projectiles, 
like two-neutron halo nuclei, are then presented and discussed.
We end this review with some views of the future in breakup-reaction theory.}

\abstract{Breakup reactions are one of the main tools 
for the study of exotic nuclei, and in particular of their continuum. 
In order to get valuable information from measurements, 
a precise reaction model coupled to a fair description 
of the projectile is needed. 
We assume that the projectile initially possesses a cluster structure, 
which is revealed by the dissociation process. 
This structure is described by a few-body Hamiltonian involving effective 
forces between the clusters. 
Within this assumption, we review various reaction models. 
In semiclassical models, the projectile-target relative motion is described 
by a classical trajectory and the reaction properties are deduced by solving 
a time-dependent \Sch equation. 
We then describe the principle and variants of the eikonal approximation: 
the dynamical eikonal approximation, 
the standard eikonal approximation, 
and a corrected version avoiding Coulomb divergence. 
Finally, we present the continuum-discretized coupled-channel method (CDCC), 
in which the \Sch equation is solved with 
the projectile continuum approximated by square-integrable states. 
These models are first illustrated by applications to two-cluster projectiles 
for studies of nuclei far from stability and of reactions useful in astrophysics. 
Recent extensions to three-cluster projectiles, 
like two-neutron halo nuclei, are then presented and discussed. 
We end this review with some views of the future in breakup-reaction theory.}

\renewcommand{\theequation}{\arabic{section}.\arabic{equation}}
\section{Introduction}
The advent of radioactive ion beams has opened a new era
in nuclear physics by providing the possibility to study nuclei
far from stability.
In particular the availability of these beams favoured the discovery 
of halo nuclei \cite{Tan85b}. 
Due to the very short lifetime of exotic nuclei, 
this study cannot be performed through usual spectroscopic techniques 
and one must resort to indirect methods. 
Breakup is one of these methods. 
In this reaction, the projectile under analysis dissociates into 
more elementary components through its interaction with a target. 
Many such experiments have been performed with the hope to probe 
exotic nuclear structures far from stability \cite{Tan96,Jon04}. 

In order to get valuable information from breakup measurements, 
one must have not only a fair description of the projectile, 
but also an accurate reaction model. 
At present, a fully microscopic description of the reaction is 
computationally unfeasible. 
Simplifying assumptions are necessary. 
First, we will discuss only elastic breakup, i.e.\ a dissociation 
process leaving the target unchanged in its ground state. 
Other channels are simulated through the use of optical potentials. 
Second, we assume a cluster structure for the projectile. 
The projectile ground state is assumed to be a bound state of the clusters 
appearing during the breakup reaction. 
The bound and continuum states of the
projectile are thus described by a few-body Hamiltonian 
involving effective forces between the constituent clusters. 
Theoretical reaction models are therefore based on this cluster description 
of the projectile and effective cluster-cluster and cluster-target interactions. 

Even within these simplifying model assumptions, a direct resolution 
of the resulting many-body \Sch equation is still not possible in most cases. 
In this article, we thus review various approximations that have been developed up to now. 

We begin with the models based on the semiclassical 
approximation \cite{AW75} in which the projectile-target 
relative motion is described by a classical trajectory. 
This approximation is valid at high energies. 
It leads to the resolution of a time-dependent \Sch equation. 
In a primary version, the time-dependent equation 
was solved at the first order of the perturbation theory \cite{AW75}. 
Then, as computers became more powerful, it could be solved 
numerically \cite{KYS94,EBB95,TW99,MB99,LSC99,CBM03c}. 
We present both versions indicating their respective advantages and drawbacks. 

We then describe the eikonal approximation \cite{Glauber} and its variants. 
The principle is to calculate the deviations from a plane-wave motion 
which are assumed to be weak at high energy.  
By comparison with the semiclassical model, it is possible to derive 
the dynamical eikonal approximation (DEA) that combines the 
advantages of both models \cite{BCG05,GBC06}. 
The standard eikonal approximation is 
obtained by making the additional adiabatic or sudden approximation, 
which neglects the excitation energies of the projectile. 
With this stronger simplifying assumption, the final state only differs 
from the initial bound state by a phase factor. 
This approach is mostly used to model reactions on light targets 
at intermediate and high energies. 
Its drawback is that the Coulomb interaction leads to a divergence 
of breakup cross sections at forward angles. 
This problem can be solved using a first-order correction of the Coulomb treatment 
within the eikonal treatment. 
A satisfactory approximation of the DEA can then be derived \cite{MBB03,AS04}: 
the Coulomb-corrected eikonal approximation (CCE), 
which remains valid for breakup on heavy targets. 
It reproduces most of the results of the DEA, 
although its computational time is significantly lower \cite{CBS08} 
which is important for the study of the breakup of three-cluster projectiles. 

Finally, we present the continuum-discretized coupled-channel method (CDCC) \cite{Kam86,TNT01}, 
in which the full projectile-target \Sch equation is solved approximately, 
by representing the continuum of the projectile with square-integrable states. 
This model leads to the numerical resolution of coupled-channel equations, 
and is suited for low- as high-energy reactions. 

All the aforementioned models have been developed initially for two-body projectiles. 
However, the physics of three-cluster systems, like two-neutron halo nuclei, 
is the focus of many experimental studies and must also be investigated 
with these models. 
We review here the various efforts that have been made in 
the past few years to extend breakup models to 
three-cluster projectiles \cite{BCD09,EMO09,RAG09}. 

In \Sec{framework}, we specify the general theoretical framework 
within which the projectile is described. 
The semiclassical model and approximate resolutions of the time-dependent 
\Sch equation are described in \Sec{tdse}. 
\Sec{eik} presents the eikonal approximation as well as the related DEA and CCE models. 
Next, in \Sec{cdcc}, the CDCC method is developed. 
In \Sec{2b}, we review applications of breakup reactions to two-body projectiles. 
In particular, we emphasize the use of breakup to study nuclei far from 
stability and as an indirect way to infer cross sections of reactions 
of astrophysical interest. 
\Sec{3b} details the recent efforts made 
to extend various reaction models to three-body projectiles. 
We end this review by presenting some views of the future in 
breakup-reaction theory. 

\section{Projectile and reaction models}
\label{framework}
\setcounter{equation}{0}
We consider the reaction of a projectile $P$ of mass $m_P$ and charge $Z_Pe$ 
impinging on a target $T$ of mass $m_T$ and charge $Z_Te$. 
The projectile is assumed to exhibit a structure made of $N$ clusters 
with masses $m_i$ and charges $Z_ie$ ($m_P = \sum_i m_i$ and $Z_P = \sum_i Z_i$). 
Its internal properties are described by a Hamiltonian $H_0$, depending on a set 
of $N-1$ internal coordinates collectively represented by notation $\xi$. 
With the aim of preserving the generality of the presentation of the reaction models, 
we do not specify here the expression of $H_0$.
Details are given in Secs.~\ref{2b} and \ref{3b},
where applications for the breakup of two- and three-body projectiles are presented.

The states of the projectile are thus described by the eigenstates of $H_0$. 
For total angular momentum $J$ and projection $M$, they are defined by
\beq
H_0 \phi^{JM}_{\tau}(E,\xi)=E \phi^{JM}_{\tau}(E,\xi),
\eeqn{e2}
where $E$ is the energy in the projectile centre-of-mass (c.m.) rest frame 
with respect to the dissociation threshold into $N$ clusters. 
Index $\tau$ symbolically represents the set of all additional quantum numbers 
that depend on the projectile structure, like spins and relative 
orbital momenta of the clusters. 
Its precise definition depends on the number of clusters and on the model 
selected when defining $H_0$. 
We assume these numbers to be discrete, though some may be continuous 
in some representations when there are more than two clusters. 
To simplify the notation, the parity $\pi$ of the eigenstates 
of $H_0$ is understood.
In the following, any sum over $J$ implicitly includes a sum over parity.

The negative-energy solutions of \Eq{e2} correspond to the 
bound states of the projectile. 
They are normed to unity. 
The positive-energy states describe the broken-up projectile 
with full account of the interactions between the clusters. 
They are orthogonal and normed according to 
$\langle\phi^{JM}_{\tau'}(E',\xi)|\phi^{JM}_\tau(E,\xi)\rangle=\delta(E-E')\delta_{\tau\tau'}$. 
To describe final states when evaluating breakup cross sections, 
we also consider the incoming scattering states $\phi^{(-)}_{\hat{k}_\xi}$. 
They correspond to positive-energy states of $H_0$
describing the $N$ clusters moving away from each other in the projectile c.m.\ frame
with specific asymptotic momenta and spin projections.
These momenta are not independent, since the sum of the asymptotic
kinetic energies of the clusters is the positive energy $E$. 
However, within that condition, their directions and, if $N > 2$,
their norms can vary.
By $\hat{k}_\xi$, we symbolically denote these directions and wave numbers, 
as well as the projections of the spins of the clusters.
These incoming scattering states are thus solutions of the \Sch equation
\beq
H_0 \phi^{(-)}_{\hat{k}_\xi}(E,\xi)=
E \phi^{(-)}_{\hat{k}_\xi}(E,\xi).
\eeqn{e2a}
They can be expanded into a linear combination of the eigenstates $\phi^{JM}_\tau$ 
of \Eq{e2} with the same energy as 
\beq
\phi^{(-)}_{\hat{k}_\xi}(E,\xi)=
\sum_{JM\tau}a^{JM}_\tau(\hat{k}_\xi)\phi^{JM}_{\tau}(E,\xi),
\eeqn{e2b}
where the coefficients $a^{JM}_\tau$ depend on the projectile structure. 
These scattering states are normed following 
$\langle\phi^{(-)}_{\hat{k}_\xi'}(E',\xi) |\phi^{(-)}_{\hat{k}_\xi}(E,\xi)\rangle 
= \delta(E-E') \delta(\hat{k}_{\xi}-\hat{k}_{\xi}')$.

The interactions between the projectile constituents and the target
are usually simulated by optical potentials chosen in the literature
or obtained by a folding procedure.
Within this framework the description of the reaction reduces to the
resolution of an $(N+1)$-body \Sch equation 
\beq
\left[ \frac{P^2}{2\mu}+H_0 +V_{PT}(\xi,\vec{R}) \right]\Psi(\xi,\vec{R})
= E_{\rm T}\Psi(\xi,\vec{R}),
\eeqn{e4}
where $\vec{R} = (R,\Omega_R) = (R,\theta_R,\varphi_R)$ 
is the coordinate of the projectile centre of mass 
relative to the target, $\vec{P}$ is the corresponding momentum, 
$\mu=m_Pm_T/(m_P+m_T)$ is the projectile-target reduced mass, 
and $E_{\rm T}$ is the total energy in the projectile-target c.m.\ frame. 
The projectile-target interaction $V_{PT}$ is expressed as 
the sum of the optical potentials (including Coulomb) that 
simulate the interactions between the projectile constituents 
and the target, 
\beq
V_{PT}(\xi,\vec{R}) = \sum_{i=1}^N V_{iT}\left(R_{iT}\right),
\eeqn{e4a}
where $\vec{R}_{iT}$ is the relative coordinate of the 
projectile cluster $i$ with respect to the target. 

The projectile being initially bound in the 
state $\phi^{J_0M_0}_{\tau_0}$ of negative energy $E_0$, 
we look for solutions of \Eq{e4} with an incoming part behaving 
asymptotically as 
\beq
\Psi(\xi,\vec{R})\flim{Z}{-\infty}e^{i\{KZ+\eta \ln[K(R-Z)]\}}
\phi^{J_0M_0}_{\tau_0}(E_0,\xi),
\eeqn{e5}
where $Z$ is the component of $\vec{R}$ in the incident-beam direction. 
The wavenumber $K$ of the projectile-target relative motion is 
related to the total energy $E_{\rm T}$ by 
\beq
E_{\rm T}=\frac{\hbar^2 K^2}{2\mu}+E_0.
\eeqn{e5a}
The $P$-$T$ Sommerfeld parameter is defined as 
\beq
\eta = Z_P Z_T e^2/\hbar v,
\eeqn{e5b}
where $v =\hbar K/\mu$ is the initial $P$-$T$ relative velocity. 

A first idea that may come to mind is to solve \Eq{e4} exactly, 
e.g., within the Faddeev framework or its extensions. 
However, the infinite range of the Coulomb interaction between 
the projectile and the target renders the standard equations ill-defined. 
Only recently significant progress has been made. 
For example, in Refs.~\cite{DFS05c,DFS05l}, 
this problem is tackled by using an appropriate screening of the Coulomb force. 
This technique has been used to successfully describe the 
elastic scattering and breakup of the deuteron on various targets. 
However, it has long been limited to light targets 
(see \Ref{DMC07} for a recent extension to a heavier target).
To obtain a model that is valid for all types of target, 
one must still resort to approximations in the resolution of \Eq{e4}. 
These approximations are made in the treatment of 
the projectile-target relative motion, 
like in the semiclassical (\Sec{tdse}) or eikonal (\Sec{eik}) approximations, 
or by using a discretized continuum, like in the CDCC method (\Sec{cdcc}). 

\section{Semiclassical approximation}
\label{tdse}
\setcounter{equation}{0}
\subsection{Time-dependent \Sch equation}
The semiclassical approximation relies on the hypothesis 
that the projectile-target relative motion can be efficiently 
described by a classical trajectory $\vec{R}(t)$ \cite{AW75}. 
It is thus valid when the de Broglie wavelength is small with respect 
to the impact parameter $b$ characterizing the trajectory, $Kb \gg 1$, 
i.e.\ when the energy is large enough. 
Along that trajectory, the projectile experiences a time-dependent 
potential $V$ that simulates the Coulomb and nuclear fields of the target. 
The internal structure of the projectile, on the contrary, 
is described quantum-mechanically by the Hamiltonian $H_0$. 
This semiclassical approximation leads to the resolution 
of the time-dependent equation 
\beq
i \hbar\frac{\partial}{\partial t}\Psi(\xi,\vec{b},t)
= \left[H_0+V(\xi,t)\right]\Psi(\xi,\vec{b},t).
\eeqn{e6}
The time-dependent potential is obtained from the difference between 
the projectile-target interaction $V_{PT}$ \eq{e4a} and 
the potential $V_{\rm traj}$ that defines the classical trajectory 
\beq
V(\xi,t) = V_{PT}[\xi,\vec{R}(t)]-V_{\rm traj}[R(t)].
\eeqn{e6b}
The potential $V_{\rm traj}$ acts as a $P$-$T$ 
scattering potential that bends the trajectory, but does not 
affect the projectile internal structure. 
Its interest lies in the fact that $V$ decreases faster than $V_{PT}$. 
Its effect amounts to changing the phase of the wave function. 
Usually it is chosen to be the Coulomb potential between the 
projectile centre of mass and the target, but it may include 
a nuclear component. 
At sufficiently high energy, the trajectory is often 
approximated by a straight line. 

For each impact parameter $b$, \Eq{e6} has to be solved 
with the initial condition that the projectile is in its ground state, 
\beq
\Psi^{(M_0)}(\xi,\vec{b},t)\flim{t}{-\infty}
\phi^{J_0 M_0}_{\tau_0}(E_0,\xi).
\eeqn{e7}
For each trajectory, the time-dependent wave function 
$\Psi^{(M_0)}$  must be calculated for the different 
possible values of $M_0$. 

\subsection{Cross sections}
From the output of the resolution of \Eq{e6}, the probability 
of being in a definite state of the projectile can 
be obtained by projecting the final wave function onto 
the corresponding eigenstate of $H_0$. 
One can for example compute the elastic scattering probability 
\beq
P_{\rm el}(b) = \frac{1}{2J_0+1}\sum_{M_0}\sum_{M_0'}
|\langle\phi^{J_0M_0'}_{\tau_0}(E_0,\xi)|
\Psi^{(M_0)}(\xi,\vec{b},t\rightarrow +\infty)\rangle|^2.
\eeqn{e7a}
This probability depends only on the norm of the impact parameter $b$
because the time-dependent wave function $\Psi^{(M_0)}$ depends on the
orientation of $\vec{b}$, i.e. on the azimuthal angle $\varphi_R$,
only through a phase that cancels out in the calculation of $P_{\rm el}$.
From this probability,
the cross section for the elastic scattering in direction $\Omega$ is obtained as 
\beq
\frac{d\sigma_{\rm el}}{d\Omega} 
= \frac{d\sigma_{\rm el}^{\rm traj}}{d\Omega} P_{\rm el}[b(\Omega)],
\eeqn{e9a}
where $b(\Omega)$ is given by the classical relation 
between the scattering angle and the impact parameter derived from potential $V_{\rm traj}$. 
The factor $d\sigma_{\rm el}^{\rm traj}/d\Omega$ is the 
elastic scattering cross section obtained from $V_{\rm traj}$. 
In most cases $d\sigma_{\rm el}^{\rm traj}/d\Omega$ is generated 
from the Coulomb interaction and is thus the $P$-$T$ Rutherford cross section. 

Likewise, a general breakup probability density can be computed 
by projecting the final wave function onto the 
ingoing scattering states of $H_0$, 
\beq
\frac{dP_{\rm bu}}{d\hat{k}_\xi dE}(b)=\frac{1}{2J_0+1}\sum_{M_0}
|\langle\phi^{(-)}_{\hat{k}_\xi}(E,\xi)|
\Psi^{(M_0)}(\xi,\vec{b},t\rightarrow +\infty)\rangle|^2.
\eeqn{e8}
After integration and summation over $\hat{k}_\xi$, the breakup probability per unit energy reads 
\beq
\frac{dP_{\rm bu}}{dE}(b)=\frac{1}{2J_0+1}\sum_{M_0}\sum_{JM\tau}
|\langle\phi^{JM}_\tau(E,\xi)|
\Psi^{(M_0)}(\xi,\vec{b},t\rightarrow +\infty)\rangle|^2.
\eeqn{e8a}
Similarly to \Eq{e9a}, a differential cross section 
for the breakup of the projectile is given by 
\beq
\frac{d\sigma_{\rm bu}}{dE d\Omega} = \frac{d\sigma_{\rm el}^{\rm traj}}{d\Omega} 
\frac{dP_{\rm bu}}{dE}[b(\Omega)].
\eeqn{e9}
The breakup cross section can then be obtained by 
summing the breakup probability over all impact parameters 
\beq
\frac{d\sigma_{\rm bu}}{dE} = 2\pi\int_{0}^{\infty} \frac{dP_{\rm bu}}{dE}(b)b db.
\eeqn{e9b}

Because of the trajectory hypothesis of the semiclassical approximation, 
the impact parameter $b$ is a classical variable. 
Therefore, no interference between the different trajectories can appear.
This is the major disadvantage of that technique since
quantal interferences can play a significant role in reactions,
in particular in those which are nuclear dominated.

\subsection{Resolution at the first order of the perturbation theory}
\label{FO}
In the early years of the semiclassical approximations, 
\Eq{e6} was solved at the first order of the perturbation theory \cite{AW75}. 
This technique, due to Alder and Winther, was applied to analyze 
the first Coulomb-breakup experiments of halo nuclei \cite{Nak94}.

The time-dependent wave function $\Psi^{(M_0)}$ is expanded upon the basis of 
eigenstates of $H_0$ in \Eq{e2}. 
At the first order of the perturbation theory, the resulting 
equation is solved by considering that $V$ is small. 
With the initial condition \eq{e7}, the wave function at first order 
is given by \cite{AW75,Ba08} 
\beq
\!\!\!\!\!e^{\frac{i}{\hbar}H_0 t} \Psi^{(M_0)}(\xi,\vec{b},t) = 
\left[1+\frac{1}{i\hbar} \int_{-\infty}^{t} \!\!\!e^{\frac{i}{\hbar}H_0 t'} 
V(\xi,t') e^{-\frac{i}{\hbar}H_0 t'} dt'\right]
\phi^{J_0M_0}_{\tau_0}(E_0,\xi).
\eeqn{e10}
Following \Eq{e8}, the general breakup probability density reads 
\beq
\!\!\!\!\!\frac{dP_{\rm bu}}{d\hat{k}_\xi dE}(b) = \frac{\hbar^{-2}}{2J_0+1}
\sum_{M_0}
\left|\int_{-\infty}^{+\infty} \!\!\!\! e^{i\omega t}
\langle\phi^{(-)}_{\hat{k}_\xi}(E,\xi)|V(\xi,t)
|\phi^{J_0M_0}_{\tau_0}(E_0,\xi)\rangle dt\right|^2,
\eeqn{e11}
where $\omega=(E-E_0)/\hbar$. 
The breakup probability per unit energy reads 
\beq
\!\!\!\!\!\frac{dP_{\rm bu}}{dE}(b) = \frac{\hbar^{-2}}{2J_0+1}
\sum_{M_0}\sum_{JM\tau}
\left|\int_{-\infty}^{+\infty} \!\!\!\! e^{i\omega t}
\langle\phi^{JM}_\tau(E,\xi)|V(\xi,t)
|\phi^{J_0M_0}_{\tau_0}(E_0,\xi)\rangle dt\right|^2\!\!\!.
\eeqn{e11aa}

With \Eq{e10}, exact expressions can be calculated when considering a purely Coulomb 
$P$-$T$ interaction for straight-line trajectories in the far-field approximation \cite{EB02}, 
i.e.\ by assuming that the charge densities 
of the projectile and target do not overlap during the collision. 
One obtains  
\beq
\lefteqn{\langle\phi^{JM}_\tau(E,\xi)|\Psi^{(M_0)}(\xi,\vec{b},t\rightarrow +\infty)\rangle =}\nonumber\\
 & & Z_T e\, \frac{e^{-iEt/\hbar}}{i\hbar}
\sum_{\lambda\mu} \frac{4\pi}{2\lambda+1}I_{\lambda\mu}(\omega,b)
\langle\phi^{JM}_\tau(E,\xi)|{\cal M}^{\rm E\lambda}_\mu(\xi)
|\phi^{J_0M_0}_{\tau_0}(E_0,\xi)\rangle,
\eeqn{e11a}
where ${\cal M}^{\rm E\lambda}_\mu$ are the electric multipoles operators of rank $\lambda$, 
and $I_{\lambda\mu}$ are time integrals (see, e.g., Eq.~(13) of \Ref{CB05}) 
that can be evaluated analytically as \cite{EB02} 
\beq
I_{\lambda \mu}(\omega,b) = \sqrt{\frac{2\lambda+1}{\pi}}\frac{1}{v} 
\frac{i^{\lambda+\mu}}{\sqrt{(\lambda+\mu)!(\lambda-\mu)!}} 
\left(-\frac{\omega}{v}\right)^\lambda K_{|\mu|}\left(\frac{\omega b}{v}\right),
\eeqn{e11b}
where $K_n$ is a modified Bessel function \cite{AS70}. 

If only the dominant dipole term E1 of the interaction is considered, 
the breakup probability \eq{e11aa} reads \cite{SLY03} 
\beq
\lefteqn{\frac{dP_{\rm bu}^{\rm E1}}{dE}(b)=\frac{16\pi}{9}
\left(\frac{Z_T e}{\hbar v}\right)^2} \nonumber\\
&\times&
\left(\frac{\omega}{v}\right)^2
\left[K_1^2\left(\frac{\omega b}{v}\right)
+K_0^2\left(\frac{\omega b}{v}\right)\right]
\frac{dB({\rm E1})}{dE}.
\eeqn{e11c}
The last factor is the dipole strength function per energy unit \cite{SLY03},
\beq
\frac{dB({\rm E1})}{dE}&=&\frac{1}{2J_0+1}\sum_{\mu M_0}
\int\!\!\!\!\!\!\!\!\!\!\sum 
d\hat{k}_\xi |\langle\phi^{(-)}_{\hat{k}_\xi}(E,\xi)|
{\cal M}^{\rm E1}_\mu(\xi)|\phi^{J_0M_0}_{\tau_0}(E_0,\xi)\rangle|^2\nonumber \\
 &=&\frac{1}{2J_0+1}\sum_{\mu M_0}\sum_{JM\tau}
|\langle\phi^{JM}_\tau(E,\xi)|
{\cal M}^{\rm E1}_\mu(\xi)|\phi^{J_0M_0}_{\tau_0}(E_0,\xi)\rangle|^2.
\eeqn{e11d}
Since modified Bessel functions decrease exponentially, the asymptotic behaviour 
of $dP_{\rm bu}^{\rm E1}/dE$ for $b \rightarrow \infty$ is proportional 
to $\exp(-2\omega b/v)$.

In the case of a purely Coulomb $P$-$T$ interaction, 
the first order of the perturbation theory exhibits many appealing aspects. 
First, it can be solved analytically. 
Second, the dynamics part ($I_{\lambda\mu}$) and structure part 
(matrix elements of ${\cal M}^{\rm E\lambda}_\mu$) 
are separated in the expression of the breakup amplitudes \eq{e11a}. 
This first-order approximation has therefore often been used to analyze Coulomb-breakup 
experiments by assuming pure E1 breakup (see, e.g., \Ref{Nak94}). 
However, as will be seen later, higher-order and nuclear-interaction effects 
are usually not negligible, and a proper analysis of experimental data 
requires a more sophisticated approximation. 

\subsection{Numerical resolution}
\label{numtdse}
The time-dependent \Sch equation can also be solved numerically.
Various groups have developed algorithms for that purpose
\cite{KYS94,EBB95,TW99,MB99,LSC99,CBM03c,KYS96,Fal02}.
They make use of an approximation of the
evolution operator $U$ applied iteratively to the initial
bound state wave function following the scheme
\beq
\Psi^{(M_0)}(\xi,\vec{b},t+\Delta t)=U(t+\Delta t,t)\Psi^{(M_0)}(\xi,\vec{b},t).
\eeqn{e12}
Although higher-order algorithms exist (see, e.g., \Ref{BGC03}), 
all practical calculations are performed with second-order
approximations of $U$.
Various expressions of this approximation exist, depending mainly
on the way of representing the time-dependent projectile wave function.
However they are in general similar to \cite{CBM03c}
\beq
U(t+\Delta t)= e^{-i\frac{\Delta t}{2\hbar}V(\xi,t+\Delta t)} 
e^{-i\frac{\Delta t}{\hbar}H_0}
e^{-i\frac{\Delta t}{2\hbar}V(\xi,t)}+{\cal O}(\Delta t^3).
\eeqn{e13}
With this expression, the time-dependent potential can be 
treated separately from the time-independent Hamiltonian $H_0$, 
which greatly simplifies the calculation of the time evolution 
when the wave functions are discretized on a mesh \cite{CBM03c}. 

The significant advantage of this technique over the
first order of perturbation is that it naturally includes
higher-order effects.
Moreover, the nuclear interaction between the projectile
and the target can be easily added in the numerical scheme \cite{TS01r}. 
However, the dynamical and structure evolutions being
now more deeply entangled, the analysis of the numerical resolution
of the \Sch equation is less straightforward than its first-order
approximation. The numerical technique is also much more time-consuming
than the perturbation one.
The first order of the perturbation theory therefore remains
a useful tool to qualitatively analyze calculations of Coulomb-dominated
reactions performed with more elaborate models.
Moreover, as will be seen in \Sec{cce}, it can be used to correct
the erroneous treatment of the Coulomb interaction within the eikonal
description of breakup reactions.

\begin{figure}
\center
\includegraphics[width=8cm]{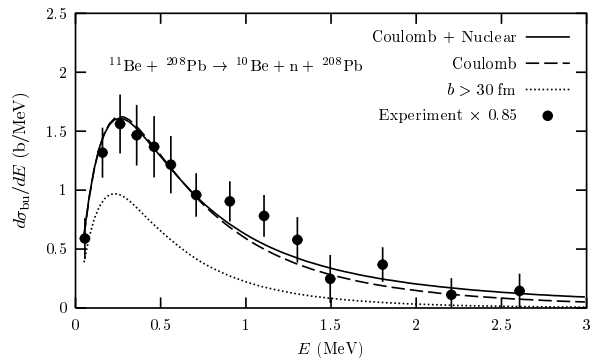}
\caption{Breakup cross section of \ex{11}Be on Pb at 68~MeV/nucleon as 
a function of the relative energy $E$ between the \ex{10}Be core and the neutron.
Calculations are performed within the semiclassical approximation
with or without nuclear interaction \cite{CBM03c}.
Experimental data \cite{Nak94} are scaled by 0.85 \cite{FNA04}.
Reprinted figure with permission from \Ref{CBM03c}.
Copyright (2003) by the American Physical Society.}
\label{f1}
\end{figure}

\fig{f1} illustrates the numerical resolution
of the time-dependent \Sch equation for the
Coulomb breakup of \ex{11}Be on lead at 68~MeV/nucleon \cite{CBM03c}.
It shows the breakup cross section as a function of
the relative energy $E$  between the \ex{10}Be core and
the halo neutron after dissociation.
The full line corresponds to the calculation with both
Coulomb and nuclear $P$-$T$ interactions.
The dashed line is the result for a purely Coulomb
potential, in which the nuclear interaction is simulated
by an impact parameter cutoff at $b_{\rm min}=13$~fm.
A calculation performed with an impact parameter cutoff at
$b_{\rm min}=30$~fm simulating a forward-angle cut
is plotted as a dotted line.
The experimental data from \Ref{Nak94} are multiplied by a factor of 
0.85 as suggested in \Ref{FNA04} after a remeasurement.

This example shows the validity of the semiclassical
approximation to describe breakup observables in the
projectile c.m.\ frame for collisions at intermediate energies.
It also confirms that for heavy targets the reaction is strongly dominated
by the Coulomb interaction. The inclusion of optical
potentials to simulate the nuclear $P$-$T$ interactions
indeed only slightly increases the breakup cross section at large energy.
This shows that Coulomb-breakup calculations are not very sensitive
to the uncertainty related to the choice of the optical potentials.
Nevertheless, since optical potentials can be very easily included
in the numerical resolution of the time-dependent \Sch equation,
they should be used so as to avoid 
the imprecise impact-parameter cutoff necessary in
purely Coulomb calculations.

\section{Eikonal approximations}
\label{eik}
\setcounter{equation}{0}
\subsection{Dynamical eikonal approximation}
\label{dea}
Let us now turn to a purely quantal treatment providing 
approximate solutions of the \Sch equation \eq{e4}. 
At sufficiently high energy, the projectile 
is only slightly deflected by the target. 
The dominant dependence of the $(N+1)$-body wave function 
$\Psi$ on the projectile-target coordinate $\vec{R}$ 
is therefore in the plane wave contributing to
the incident relative motion \eq{e5}. 
The main idea of the eikonal approximation 
is to factorize that plane wave out of the wave function 
to define a new function $\widehat{\Psi}$ 
whose variation with $\vec{R}$ is expected to be small 
\cite{Glauber,SLY03,BD04} 
\beq
\Psi(\xi,\vec{R})=e^{iKZ}\widehat\Psi(\xi,\vec{R}).
\eeqn{e20}
With factorization \eq{e20} and energy conservation \eq{e5a}, 
the \Sch equation \eq{e4} becomes 
\beq
\left[ \frac{P^2}{2\mu}+v P_Z+H_0-E_0+V_{PT}(\xi,\vec{R})\right]
\widehat{\Psi}(\xi,\vec{R}) = 0,
\eeqn{e22}
where the relative velocity $v$ between projectile and target 
is assumed to be large. 

The first step in the eikonal approximation is to assume the
second-order derivative $P^2/2\mu$
negligible with respect to the first-order derivative $vP_Z$, 
\beq
\frac{P^2}{2\mu} \widehat{\Psi}(\xi,\vec{R})\ll
vP_Z\widehat{\Psi}(\xi,\vec{R}).
\eeqn{e22b}
This first step leads to the second-order equation (but now first-order in $Z$),
\beq
i\hbar v \frac{\partial}{\partial Z}\widehat\Psi(\xi,\vec{b},Z)=
\left[H_0-E_0+V_{PT}(\xi,\vec{R})\right]
\widehat{\Psi}(\xi,\vec{b},Z),
\eeqn{e23}
where the dependence of the wave function on the 
longitudinal $Z$ and transverse $\vec{b}$ parts of the projectile-target 
coordinate $\vec{R}$ has been made explicit. 
This equation is mathematically equivalent to the time-dependent 
\Sch equation \eq{e6} for straight-line trajectories 
with $t$ replaced by $Z/v$.
It can thus be solved using any of the algorithms cited in \Sec{numtdse}. 
However, contrary to time-dependent models, it is obtained
without the semiclassical approximation.
The projectile-target coordinate components $\vec{b}$ and $Z$ are
thus quantal variables.
Interferences between solutions obtained at different $b$ values are thus
taken here into account.
This first step is known as the dynamical eikonal approximation (DEA) \cite{BCG05,GBC06}. 

\subsection{Cross sections}
\label{sec_eik}
 
The transition matrix element for elastic scattering into direction 
$\Omega = (\theta,\varphi)$ of the final momentum $\vec{K} = (K,\Omega)$ 
of the projectile in the c.m.\ frame reads \cite{Aus70} 
\beq
T_{fi} = \la e^{i\vec{K}\cdot\vec{R}} \phi^{J_0M'_0}_{\tau_0}(E_0,\xi) 
| V_{PT}(\xi,\vec{R}) | \Psi^{(M_0)}(\xi,\vec{R}) \ra,
\eeqn{s1}
where $\Psi^{(M_0)}$ is the exact solution of the \Sch equation \eq{e4} 
with the asymptotic condition \eq{e5}.
By using Eqs.~\eq{e20}, \eq{e2}, and \eq{e23}, one obtains the approximation \cite{BCG05} 
\beq
T_{fi} &=& \la e^{i\vec{K}\cdot\vec{R}} \phi^{J_0M'_0}_{\tau_0}(E_0,\xi) | 
e^{iKZ} \left[ H_0 - E_0 + V_{PT}(\xi,\vec{R}) \right] | \widehat{\Psi}^{(M_0)}(\xi,\vec{R}) \ra 
\nonumber \\
&\approx & i\hbar v \int d\vec{R}\, e^{-i \vec{q} \cdot \vec{b}} 
\frac{\partial}{\partial Z} \la \phi^{J_0M_0'}_{\tau_0}(E_0,\xi) | \widehat{\Psi}^{(M_0)}(\xi,\vec{R}) \ra,
\eeqn{s2}
where the transfered momentum $\vec{q} = \vec{K} - K\hat{\vec{Z}}$ is assumed to be purely transverse, 
i.e.\ $\exp[i(\vec{K} \cdot \hat{\vec{Z}}-K)]$, is neglected. 
The norm of $q$ is linked to the scattering angle by 
\beq
q=2K\sin\theta/2.
\eeqn{s2b}
Let us define the elastic amplitude 
\beq
S_{{\rm el},M'_0}^{(M_0)} (\vec{b}) = \la \phi^{J_0M'_0}_{\tau_0}(E_0,\xi) | 
\widehat{\Psi}^{(M_0)}(\xi,\vec{b},Z\rightarrow +\infty) \ra - \delta_{M'_0 M_0}.
\eeqn{s3}
The transition matrix element \eq{s2} reads after integration over $Z$, 
\beq
T_{fi} = i\hbar v \int d\vec{b} e^{-i \vec{q} \cdot \vec{b}} e^{i(M_0-M'_0)\varphi_R} 
S_{{\rm el},M'_0}^{(M_0)} (b\hat{\vec{X}}),
\eeqn{s4}
where $\varphi_R$ is the azimuthal angle characterizing $\vec{b}$. 
The phase factor $\exp[i(M_0-M'_0)\varphi_R]$ arises from the rotation of the wave functions 
when the orientation of $\vec{b}$ varies \cite{GBC06}. 
The integral over $\varphi_R$ can be performed analytically,
which leads to the following expression for the
elastic differential cross section \cite{GBC06}
\beq
\frac{d\sigma_{\rm el}}{d\Omega} &=& K^2 \frac{1}{2J_0+1} \sum_{M_0M'_0} 
\left| \int_0^{\infty} b db J_{|M_0 - M'_0|} (qb) S_{{\rm el},M'_0}^{(M_0)} (b\hat{\vec{X}}) \right|^2,
\eeqn{s6}
where $J_m$ is a Bessel function \cite{AS70}. 
From \Eq{s6}, one can see that contrary to the semiclassical approximation \eq{e9a}, 
the eikonal elastic cross section is obtained as a coherent sum of elastic amplitudes over all $b$ values. 
This illustrates that quantum interferences are taken into account in the eikonal framework. 

The transition matrix element for dissociation reads
\beq
T_{fi} = \la e^{i\vec{K'}\cdot\vec{R}} \phi^{(-)}_{\hat{k}_\xi}(E,\xi)| 
V_{PT}(\xi,\vec{R})| \Psi^{(M_0)}(\xi,\vec{R}) \ra,
\eeqn{s7}
where $\vec{K'}=(K',\Omega)$ is
the final projectile-target wave vector.
One can then proceed as for the elastic scattering. 
Using Eqs.~\eq{e20}, \eq{e2a}, and \eq{e23}, 
taking into account the energy conservation, 
\beq
\frac{\hbar^2 K^2}{2\mu} + E_0 = \frac{\hbar^2 K'^2}{2\mu} + E,
\eeqn{s11}
and assuming the transfered momentum $\vec{q}=\vec{K'}-K\vec{\hat{Z}}$ 
to be purely transverse, the transition matrix element is expressed as 
\beq
T_{fi} \approx i\hbar v \int d\vec{b}  e^{-i\vec{q} \cdot \vec{b}}
S_{\rm bu}^{(M_0)}(E,\hat{k}_\xi,\vec{b}),
\eeqn{s13}
with the breakup amplitude 
\beq
S_{\rm bu}^{(M_0)}(E,\hat{k}_\xi,\vec{b}) =
\la \phi^{(-)}_{\hat{k}_\xi}(E,\xi) | 
\widehat{\Psi}^{(M_0)}(\xi,\vec{b},Z\rightarrow+\infty) \ra.
\eeqn{s14}
The differential cross section for breakup is given by
\beq
\frac{d\sigma}{d\hat{k}_\xi dE d\Omega} \propto \frac{1}{2J_0+1} \sum_{M_0} 
\left|\int d\vec{b}  e^{-i\vec{q} \cdot \vec{b}} S^{(M_0)}_{\rm bu}(E,\hat{k}_\xi,\vec{b})\right|^2,
\eeqn{s16}
where the proportionality factor depends on the phase space. 
Like the elastic scattering cross section \eq{s6}, it is obtained 
from a coherent sum of breakup amplitudes \eq{s14}, 
confirming the quantum-mechanical character of the eikonal approximation. 
Here also, the integral over $\varphi_R$ can be performed analytically 
and leads to Bessel functions \cite{GBC06}.

By integrating expression \eq{s16} over unmeasured quantities,
one can obtain
the breakup cross sections with respect to the desired variables,
like the internal excitation energy of the projectile.
Since these operations depend on the projectile internal structure,
we delay the presentation of some detailed expressions to 
Secs.~\ref{2b} and \ref{3b} treating of two-body \cite{GBC06} 
and three-body \cite{BCD09} breakup.

\subsection{Standard eikonal approximation}
In most references, the concept of eikonal approximation 
involves a further simplification to the DEA \cite{HBE96,SLY03}. 
This adiabatic, or sudden, approximation consists in neglecting 
the excitation energy of the projectile compared 
to the incident kinetic energy. 
It comes down to assume the low-lying spectrum of the projectile
to be degenerate with its ground state, i.e.\ to consider the internal
coordinates  of the projectile as frozen during the reaction \cite{SLY03}.
This approximation therefore holds only for high-energy collisions
that occur during a very brief time.
This second assumption 
leads to neglect the term $H_0-E_0$ in the DEA equation \eq{e23} which then reads 
\beq
i\hbar v \frac{\partial}{\partial Z}\widehat\Psi(\xi,\vec{b},Z)=
V_{PT}(\xi,\vec{R})\widehat{\Psi}(\xi,\vec{b},Z).
\eeqn{e24}
The solution of \Eq{e24} that follows the asymptotic condition \eq{e5} 
exhibits the well-known eikonal form \cite{Glauber,BD04} 
\beq
\widehat\Psi^{(M_0)}(\xi,\vec{b},Z)=
\exp\left[-\frac{i}{\hbar v}
\int_{-\infty}^Z V_{PT}(\xi,\vec{b},Z')dZ'\right]
\phi^{J_0M_0}_{\tau_0}(E_0,\xi).
\eeqn{e26}
After the collision, the whole information about the change in the 
projectile wave function is thus contained in the phase shift 
\beq
\chi(\vec{s}_\xi,\vec{b})=-\frac{1}{\hbar v}\int_{-\infty}^{+\infty}
V_{PT}(\xi,\vec{R})dZ.
\eeqn{e27}
Due to translation invariance, this eikonal phase $\chi$ depends only
on the transverse components $\vec{b}$ 
of the projectile-target coordinate $\vec{R}$ 
and $\vec{s}_\xi$ of the projectile internal coordinates $\xi$. 
Cross sections within this standard eikonal approximation are obtained
as explained in \Sec{sec_eik}, replacing $\widehat\Psi^{(M_0)}$ by
$e^{i\chi}\phi^{J_0M_0}_{\tau_0}$.

Being obtained from the adiabatic approximation, 
expressions \eq{e26} and \eq{e27} are valid only for short-range potentials. 
For the Coulomb interaction, 
the assumption that the reaction takes place in a short time no longer holds,
due to its infinite range. 
The adiabatic approximation thus fails 
for Coulomb-dominated reactions \cite{SLY03}. 
Besides imprecise uses of a cutoff at large impact parameters \cite{AS00}, 
there are two ways to avoid this problem. 
The first is not to make the adiabatic approximation, 
i.e.\ to resort to the more complicated DEA (see \Sec{dea}). 
The second is to correct the eikonal phase 
for the Coulomb interaction as suggested in \Ref{MBB03} (see \Sec{cce}). 
Nevertheless, as shown in \Ref{GBC06}, the Coulomb divergence does 
not affect eikonal calculations performed on light targets 
at high enough energies. 
Most of the nuclear-dominated reactions can thus be 
analyzed within an eikonal model including 
the adiabatic approximation (see, e.g., \Ref{HT03}). 

\begin{figure}
\center
\includegraphics[width=8cm]{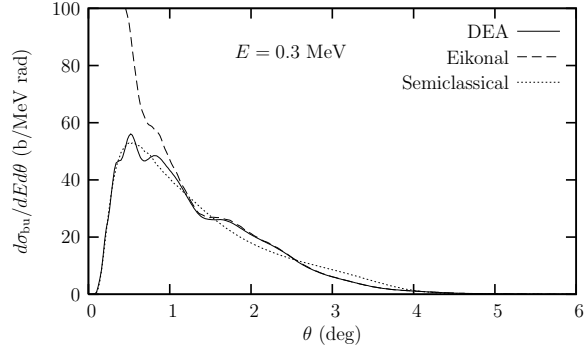}
\caption{Breakup cross section of $^{11}$Be on Pb at 69~MeV/nucleon as 
a function of the $P$-$T$ scattering angle in the $P$-$T$ c.m.\ frame
for a \ex{10}Be-n energy $E=0.3$~MeV. 
Calculations are performed within the DEA, usual eikonal, 
and semiclassical approximations \cite{phdGG}. }
\label{f2}
\end{figure}

\fig{f2} illustrates the difference between the DEA (full line), 
the usual eikonal approximation (dashed line) 
and the semiclassical approximation (dotted line) 
when Coulomb dominates. 
It shows the breakup cross section of \ex{11}Be on Pb 
at 69 MeV/nucleon for a \ex{10}Be-n relative energy of 0.3~MeV 
as a function of the $P$-$T$ scattering angle. 
As explained above, the usual eikonal approximation diverges for 
the Coulomb-dominated breakup, i.e.\ at forward angles. 
The DEA, which does not include the adiabatic approximation, 
exhibits a regular behaviour at these angles. 
Interestingly, the semiclassical approximation follows 
the general behaviour of the DEA, except for the oscillations 
due to quantum interferences between different $b$ values. 
The DEA has therefore the advantage of being valid for 
describing any breakup observable on both light and heavy targets. 

The nuclei studied through breakup reactions being exotic,
it may be difficult, if not impossible, to find
optical potentials that describe the scattering of the clusters
by the target. 
One way to circumvent that problem
is to resort to what is usually known as the
Glauber model \cite{GM70,HBE96,SLY03,BD04}. 
This model has been mostly used to calculate 
total and reaction cross sections. 
At the optical-limit approximation (OLA) of the Glauber model,
correlations in the cluster and target wave functions
are neglected. 
The nuclear component of the eikonal
phase shift for cluster $i$ is then expressed as a function 
of the densities $\rho_T$ of the target and $\rho_i$ of the cluster,
and of a profile function $1-e^{i\chi_{\rm NN}}$ 
that corresponds to an effective nucleon-nucleon interaction. 
The nuclear component of the eikonal phase shift is approximated by \cite{SLY03} 
\beq
\chi^N_i(\vec{b}_i)=i\int\!\!\!\!\int\rho_T(\vec{r}_T)\rho_i(\vec{r}_i) 
[1-e^{i\chi_{\rm NN}(|\vec{b}_i-\vec{s}_T+\vec{s}_i|)}] d\vec{r}_Td\vec{r}_i,
\eeqn{e28}
where $\vec{s}_T$ and $\vec{s}_i$ are the transverse components 
of the internal coordinates $\vec{r}_T$ of the target 
and $\vec{r}_i$ of cluster $i$, respectively, and $\vec{b}_i$ is 
the transverse component of the c.m.\ coordinate of cluster $i$.
The OLA is therefore equivalent to the
double-folding of an effective nucleon-nucleon interaction.
The density of the target can usually be obtained from
experimental data. The cluster density being unknown,
it has to be estimated from some structure model, like
a mean-field calculation.
The profile function is usually parametrized as \cite{SLY03,AHK08}
\beq
1-e^{i\chi_{\rm NN}(b)}=\frac{1-i\alpha_{\rm NN}}{4\pi\beta_{\rm NN}}
\sigma^{\rm tot}_{\rm NN} \exp\left(-\frac{b^2}{2\beta_{\rm NN}}\right),
\eeqn{e35}
where $\sigma^{\rm tot}_{\rm NN}$ is the total cross section
for the N-N collision, $\alpha_{\rm NN}$ is the ratio of the real part 
to the imaginary part of the N-N scattering amplitude,
and $\beta_{\rm NN}$ is the slope parameter of the N-N elastic
differential cross section.
These parameters depend on the nucleon type
(p or n) and on the incident energy.
Their values can be found in the literature (see, e.g., \Ref{AHK08}). 
The validity of the Glauber approximation is discussed in \Ref{YMO08}. 

\subsection{Coulomb-corrected eikonal approximation}
\label{cce}
The eikonal approximation gives excellent results for nuclear-dominated 
reactions \cite{SLY03,GBC06}. 
However, as mentioned above, it suffers from 
a divergence problem when the Coulomb interaction becomes significant. 
To explain this, let us divide the eikonal phase \eq{e27} 
into its Coulomb and nuclear contributions 
\beq
\chi(\vec{s}_\xi,\vec{b})=\chi_{PT}^C(b)+\chi^C(\vec{s}_\xi,\vec{b})+\chi^N(\vec{s}_\xi,\vec{b}).
\eeqn{e40}
In this expression, $\chi^C_{PT}$ is the global elastic Coulomb eikonal phase 
between the projectile and the target. 
However, Coulomb forces not only act globally on the projectile, 
they also induce `tidal' effects due to their different actions on the various clusters. 
The tidal Coulomb phase $\chi^C$ is due to the difference between the cluster-target
and projectile-target bare Coulomb interactions. 
The remaining phase $\chi^N$ contains effects of the nuclear forces as well as 
of differences between Coulomb forces taking  the finite size of the
clusters into account 
and the bare Coulomb forces. 

At the eikonal approximation, the integral \eq{e27} defining $\chi^C_{PT}$
diverges and must be calculated  with a cutoff \cite{Glauber,SLY03}.
Up to an additional cutoff-dependent term that plays no role
in the cross sections, it can be written as \cite{BD04}
\beq
\chi^C_{PT}(b) = 2\eta\ln(K b),
\eeqn{e41}
where appears the projectile-target Sommerfeld parameter $\eta$ defined in \Eq{e5b}. 
The phase \eq{e41} depends only on $b$. 

The tidal Coulomb phase is computed with \Eq{e27} 
for the difference between the bare Coulomb interactions 
for the clusters in the projectile and the global 
$P$-$T$ Coulomb interaction, 
\beq
\chi^C(\vec{s}_\xi,\vec{b}) &=& - \frac{\eta}{Z_P} \int_{-\infty}^{+\infty}
\left(\sum_{i=1}^N \frac{Z_i}{|\vec{R}_{iT}|} - \frac{Z_P}{|\vec{R}|}\right) dZ.
\eeqn{e41a}
It can be expressed analytically. 
Because of the long range of the E1 component of the Coulomb force, 
this phase behaves as $1/b$ at large distances \cite{GBC06,CBS08}. 
In the calculation of the breakup cross sections \eq{s16}, 
the integration over $bdb$ diverges for small $q$ values, 
i.e. at forward angles,  
because of the corresponding $1/b$ asymptotic behaviour of the breakup amplitude, 
as illustrated in \fig{f2}. 
This divergence occurs only in the first-order term $i\chi^C$ 
of the expansion of the eikonal Coulomb amplitude $\exp(i \chi^C)$. 

As seen in \Sec{FO}, the first order approximation \eq{e11c} 
decreases exponentially at large $b$ and hence does not display such a divergence. 
A plausible correction is therefore to replace the
exponential of the eikonal phase according to \cite{MBB03,AS04}
\beq
e^{i\chi} \rightarrow e^{i\chi_{PT}^C} \left(e^{i\chi^C}-i\chi^C+i\chi^{FO}\right) e^{i\chi^N},
\eeqn{e42}
where $\chi^{FO}$ is the result of first-order perturbation theory \eq{e10},
\beq
\chi^{FO}(\xi,\vec{b})= - \frac{\eta}{Z_P}
\int_{-\infty}^{+\infty} e^{i\omega Z/v}
\left(\sum_{i=1}^N \frac{Z_i}{|\vec{R}_{iT}|} - \frac{Z_P}{|\vec{R}|}\right)dZ.
\eeqn{e42a}
Note that because of the phase $e^{i\omega Z/v}$,
the integrand in \Eq{e42a} does not exhibit a translational invariance.
The first-order phase $\chi^{FO}$ depends on all internal coordinates
of the projectile.
When the adiabatic approximation is applied to \Eq{e42a},
i.e.\ when $\omega$ is set to 0, one recovers exactly
the Coulomb eikonal phase \eq{e41a}.
This suggests that without adiabatic approximation the
first-order term of $\exp(i\chi^C)$ would be $i\chi^{FO}$ \eq{e42a} instead
of $i\chi^C$ \eq{e41a}, intuitively validating the correction \eq{e42}.
Furthermore, since a simple analytic expression is available for each of the Coulomb 
multipoles (see \Sec{FO}), this correction is easy to implement.

With this Coulomb correction, the breakup of loosely-bound projectiles can
be described within the eikonal approximation
taking on (nearly) the same footing both Coulomb and nuclear
interactions at all orders.
This approximation has been tested and validated
for a two-body projectile in Ref.~\cite{CBS08}.
Note that in all practical cases \cite{AS04,CBS08,BCD09},
only the dipole term of the first-order expansion \eq{e11a}
is retained to evaluate $\chi^{FO}$

\begin{figure}
\center
\includegraphics[width=8cm]{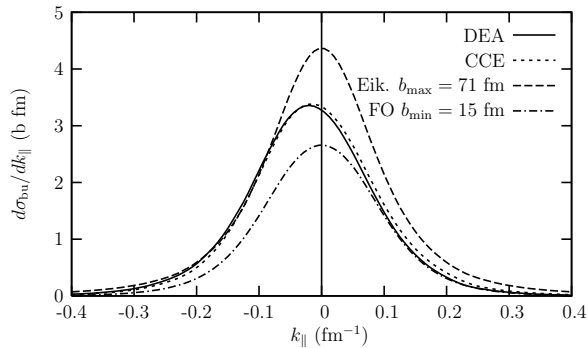}
\caption{Breakup of \ex{11}Be on Pb at 69~MeV/nucleon.
The parallel-momentum distribution between the \ex{10}Be core
and the halo neutron is computed within the DEA, the CCE,
the eikonal approximation including the adiabatic approximation,
and the first-order of the perturbation theory \cite{CBS08}.
Reprinted figure with permission from \Ref{CBS08}.
Copyright (2008) by the American Physical Society.}
\label{f3}
\end{figure}

\fig{f3} illustrates the accuracy of the CCE for the breakup
of \ex{11}Be on lead at 69~MeV/nucleon \cite{CBS08}.
The figure presents the parallel-momentum distribution
between the \ex{10}Be core and the halo neutron after dissociation.
This observable has been computed within the DEA (full line), which
serves as a reference calculation, the CCE (dotted line),
the eikonal approximation including the adiabatic approximation (dashed line),
and the first-order of the perturbation theory (dash-dotted line).
The usual eikonal approximation requires a cutoff at large impact parameter
to avoid divergence. The value $b_{\rm max}=71$~fm is chosen
from the value prescribed in \Ref{AS00}.
At the first order or the perturbation theory, the nuclear interaction
is simulated by an impact parameter cutoff at $b_{\rm min}=15$~fm.

We first see that the magnitude of the CCE cross section 
is close to the DEA one, whereas, the other two approximations 
give too large (eikonal) or too small (first order) cross sections. 
Moreover the CCE reproduces nearly perfectly the shape 
of the DEA distribution. In particular the asymmetry, 
due to dynamical effects, is well reproduced. 
This result suggests that in addition to solving the 
Coulomb divergence problem introduced by the adiabatic approximation, 
the CCE also restores some dynamical and higher-order effects 
missing in its ingredients, the usual eikonal approximation and the first order 
of the perturbation theory.

\section{Continuum-discretized coupled-channel method}
\label{cdcc}
\setcounter{equation}{0}

The CDCC method is a fully quantal approximation which does not imply some restriction 
on energies. 
Its main interest lies in low energies where the previous methods are not valid. 
The principle of the CDCC method is to determine, as accurately as possible, the 
scattering and dissociation cross sections of a nucleus with a simplified 
treatment of the final projectile continuum states. 
To this end, these states describing the relative motions of the unbound fragments 
are approximately described by square-integrable wave functions at discrete energies. 
The relative motion between the projectile and target and various cross sections 
can then be obtained by solving a system of coupled-channel equations. 
The number of these equations and hence the difficulty of the numerical treatment 
increase with increasing energy. 

The CDCC method was suggested by Rawitscher \cite{Ra74} and first applied to deuteron + nucleus 
elastic scattering and breakup reactions. 
It was then extensively developed and used by several groups \cite{Kam86,AIK87,NT99,MAG01,RKM08,MAG09,DBD10}. 
Its interest has been revived by the availability of radioactive beams of weakly 
bound nuclei dissociating into two \cite{TNT01,NT99,MAG01,RKM08,MAG09,DBD10} 
or three \cite{MHO04,RAG08,EMO09,RAG09} fragments. 

We assume that the breakup process leads to $N$ clusters and 
that the cluster-target interactions do not depend on the target spin. 
The projectile wave functions $\phi^{JM}_{\tau}(E^J_{\tau B},\xi)$ describing $N$-body bound states 
at negative energies $E^J_{\tau B}$ and $\phi^{JM}_\tau(E,\xi)$ describing $N$-body scattering states 
at positive energies $E$ are defined with \Eq{e2}. 
Since the total angular momentum of the projectile-target system is a good quantum number, 
the first step consists in determining partial waves 
of the $(N+1)$-body Hamiltonian \eq{e4}. 
The general partial wave function for a total angular momentum $J_{\rm T}$ 
can be expanded over the projectile eigenstates as 
\beq
\Psi^{J_{\rm T}M_{\rm T}}(\vec{R},\xi) 
&=& \sum_{LJ\tau} \sum_B[\phi^{J}_{\tau}(E^J_{\tau B},\xi) \otimes \psi^L_{J\tau B}(\vec{R})]^{J_{\rm T}M_{\rm T}} 
\eol
& & + \sum_{LJ\tau} \int_0^\infty [\phi^{J}_\tau(E,\xi) \otimes \psi^L_{J\tau E}(\vec{R})]^{J_{\rm T}M_{\rm T}} dE.
\eeqn{x0}
In this expansion, index $B$ runs over the bound states of the projectile. 
The total angular momentum $J_{\rm T}$ results from the coupling of the orbital momentum $L$ 
of the projectile-target relative motion with the total angular momentum $J$ of the projectile state. 
The relative-motion partial waves $\psi^L_{J\tau B}$ and $\psi^L_{J\tau E}$ are unknown and must be determined. 
The parity is given by the product of $(-1)^L$ and the parity of $\phi^{JM}_{\tau}$. 
The first term of \Eq{x0} represents the elastic and inelastic channels while the second term is associated 
with the breakup contribution. 
However, the presence of the continuum renders this expression intractable. 

The basic idea of the CDCC method is to replace wave function \eq{x0} by 
\beq
\Psi^{J_{\rm T}M_{\rm T}}(\vec{R},\xi) = 
\sum_{LJ\tau n} [\phi^{J}_{\tau n}(\xi) \otimes \psi^L_{J\tau n}(\vec{R})]^{J_{\rm T}M_{\rm T}},
\eeqn{x1}
where the functions $\phi^{JM}_{\tau n}(\xi)\equiv \phi^{JM}_{\tau}(E^J_{\tau n},\xi)$ 
represent either bound states ($E^J_{\tau B}<0$) or square-integrable 
approximations of continuum wave functions ($E^J_{\tau n}>0$) at discrete energies 
\beq
E^J_{\tau n} = \langle \phi^{JM}_{\tau n}(\xi) |H_0| \phi^{JM}_{\tau n}(\xi) \rangle.
\eeqn{x2a}
Approximation \eq{x1} resembles usual coupled-channel expansions and can be treated in a similar way. 

In practice, two methods are available to perform the continuum discretization.
In the ``pseudostate'' approach,
the \Sch equation \eq{e2} is solved approximately 
by diagonalizing the projectile Hamiltonian $H_0$ either within a finite basis
of square-integrable functions or in a finite region of space. 
In both cases, square-integrable pseudostates $\phi^{JM}_{\tau n}$ are obtained. 
This approach is simple but there is little control on the obtained energies $E^J_{\tau n}$. 
Therefore, it is customary to keep only the pseudostates 
with energies below some limit $E_{\rm max}$. 

The alternative is to separate the integral over $E$ in \Eq{x0} into a limited number 
of small intervals, or ``bins'', $[E_{n-1},E_n]$ which may depend on $J$ 
and to use in each of them some average of the exact scattering states 
in this range of energies \cite{Kam86,Ra74,AIK87,NT99}. 
This ``bin'' method provides the square-integrable basis functions 
\beq
\phi^{JM}_{\tau n}(\xi) = \frac{1}{W_n} \int_{E_{n-1}}^{E_n}
\phi^{JM}_{\tau}(E,\xi) f_n(E) dE,
\eeqn{x3}
where the weight functions $f_n$ may also depend on $J$. 
Such states are orthogonal because of the orthogonality of the scattering states 
and they are normed if $W_n$ is the norm of $f_n$ over $[E_{n-1},E_n]$. 
Using \Eq{x3}, their energy \eq{x2a} is given by 
\beq
E^J_{\tau n} = \frac{1}{W_n^2} \int_{E_{n-1}}^{E_n} |f_n(E)|^2 E dE.
\eeqn{x4}
Here also, a maximum energy $E_{\rm max} \equiv E_{n_{\rm max}}$ is chosen. 
In practice, these basis states are usually constructed by averaging the scattering
states $\tilde{\phi}^{JM}_{\tau}(k,\xi)$ normalized over the wave number $k$,
often within equal momentum intervals \cite{TNT01}. 

The total wave function \eq{x1} can be rewritten as 
\beq
\Psi^{J_{\rm T}M_{\rm T}}(\vec{R},\xi) = R^{-1} \sum_{c} \Phi_{c}^{J_{\rm T}M_{\rm T}}(\Omega_R,\xi) u^{J_{\rm T}}_{c}(R),
\eeqn{x5}
where $c$ represents the channel $LJ\tau n$ and 
\beq
\Phi_{c}^{J_{\rm T}M_{\rm T}}(\Omega_R,\xi) = i^{L} 
\left[\phi^{J}_{\tau n}(\xi) \otimes Y_L(\Omega_R) \right]^{J_{\rm T}M_{\rm T}}.
\eeqn{x6}
By inserting expansion \eq{x5} in the \Sch equation \eq{e4} and using \Eq{x2a}, 
the relative wave functions $u^{J_{\rm T}}_{c}$ are given by a set of coupled equations 
\beq
\left[ -\frac{\hbar^2}{2\mu}\left( \frac{d^2}{dR^2}-\frac{L(L+1)}{R^2}\right) +E_{c}-E_{\rm T} \right] 
u^{J_{\rm T}}_{c}(R) 
+\sum_{c'} V^{J_{\rm T}}_{c,c'}(R) u^{J_{\rm T}}_{c'}(R)=0,
\eeqn{x7}
where $E_c \equiv E^{J}_{\tau n}$.
The sum over $L$ is truncated at some value $L_{\rm max}$. 
The sum over the pseudo-states or bins is limited by the selected maximum energy $E_{\rm max}$. 
The CDCC problem is therefore equivalent to a system of coupled equations 
where the potentials are given by
\beq
V^{J_{\rm T}}_{c,c'}(R) = 
\langle \Phi_{c}^{J_{\rm T}M_{\rm T}}(\Omega_R,\xi) | V_{PT}(\vec{R},\xi) |
\Phi_{c'}^{J_{\rm T}M_{\rm T}}(\Omega_R,\xi) \rangle.
\eeqn{x8}
This matrix element involves a multidimensional integral over $\Omega_R$ 
and over the internal coordinates $\xi$. 
In general, the potentials are expanded into multipoles corresponding to the 
total angular momentum operator $\vec{J}_{\rm T}$ of the system. 
This may allow an analytical treatment of angular integrals. 

System \eq{x7} must be solved with the boundary condition 
\beq
u^{J_{\rm T}}_{c}(R)\flim{R}{\infty} v_c^{-1/2} 
\left[I_c(K_c R)\delta_{cc_0}-O_c(K_c R)S^{J_{\rm T}}_{cc_0}\right],
\eeqn{x8a}
where $c_0$ is the incoming channel. 
The asymptotic momentum in channel $c$ reads 
\beq
K_c = \sqrt{2\mu(E_{\rm T}-E_c)/\hbar^2},
\eeqn{x8b}
and $v_c=\hbar K_c/\mu$ is the corresponding velocity.
In \Eq{x8a}, $I_c=G_c-iF_c$ and $O_c=I_c^*$ are the incoming and outgoing
Coulomb functions, respectively \cite{AS70}, and 
the element $S^{J_{\rm T}}_{cc_0}$ of the collision matrix is 
the amplitude for populating channel $c$ from initial channel $c_0$. 

Various methods have been developed to solve system \eq{x7}
(see, e.g., \Ref{Th88}). 
A convenient approach is the $R$-matrix formalism \cite{DB10}, 
which is both simple and accurate. 
The configuration space is divided into two regions: the internal ($R<a$) and external 
($R>a$) regions, where $a$ is the channel radius. 
In the external region, the potential matrix defined by \Eq{x8} 
can be well approximated by its diagonal Coulomb asymptotic form. 
Hence the wave function is replaced by combinations of Coulomb functions. 
In the internal region, the radial wave functions $u^{J_{\rm T}}_{c}$ can be expanded 
over some basis \cite{DB10}. 
A significant simplification occurs when using Lagrange functions \cite{BH86,Ba06,DBD10}. 

A scattering wave function verifying the initial condition \eq{e5} 
is then constructed with the different partial waves. 
Inserting this CDCC approximate wave function in \Eq{s7} enables calculating
transition matrix elements towards pseudostates or bin states as a function of
the collision matrices $S^{J_{\rm T}}$ (see Eq.~(5) of \Ref{TNT01}).
Since these transition matrix elements are obtained only at discrete energies
$E^J_{n\tau}$, they must be interpolated in order to obtain breakup cross
sections at all energies. 

\begin{figure}[t]
\center
\includegraphics[width=7cm]{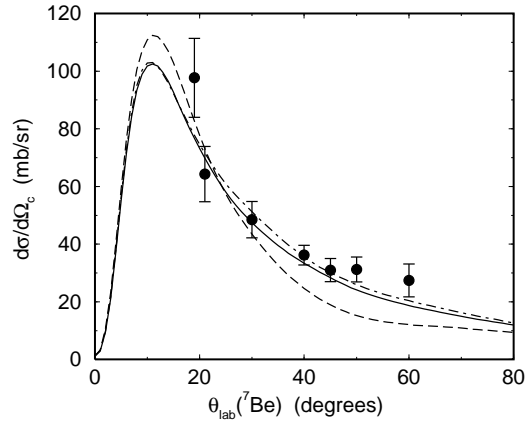}
\caption{$^7$Be angular distribution after the breakup of $^8$B on $^{58}$Ni 
at 25.8 MeV computed within a CDCC model  \cite{TNT01}. 
The convergence of the numerical scheme is illustrated 
with various maximum values of the $^7$Be-p relative orbital 
momentum $l$ in the continuum 
and various maximum values of multipole order $\lambda$ of the potential expansion in \Eq{x8}: 
$l \le 3$, $\lambda \le 2$ (dashed line),
$l \le 3$, $\lambda \le 3$ (full line),
$l \le 4$, $\lambda \le 4$ (dash-dotted line).
Experimental data from \Ref{KGP01}.
Reprinted figure with permission from \Ref{TNT01}.
Copyright (2001) by the American Physical Society.}
\label{fcdcc}
\end{figure}

The CDCC method has first been applied to two-body projectiles. 
As an example, \fig{fcdcc} shows the convergence 
of the breakup of $^8$B on $^{58}$Ni at 25.8 MeV. 
The convergence concerns the set of partial waves $l$ in the $^7$Be-p continuum 
of the projectile and the number of multipoles in the expansion 
of the potential appearing in matrix elements \eq{x8}. 
The validity of CDCC has been tested for breakup observables 
in a comparison with three-body Faddeev calculations \cite{DMC07}. 
The agreement between both sets of results is good except 
when the coupling with the transfer channel is important.

Let us also mention extensions beyond the simple two-body model of the projectile 
by allowing the core to be in an excited state \cite{SNT06r,SNT06}. 
These references present total cross sections for the breakup on a $^9$Be target 
of $^{11}$Be into $^{10}$Be + n and of $^{15}$C into $^{14}$C + n 
calculated by including core deformations.
This extension of CDCC known as XCDCC leads to very long computational times.

The extension of CDCC to three-body projectiles is more recent 
\cite{MHO04,MEO06,RAG08,EMO09,RAG09}. 
The calculations are still much more time-consuming since the projectile wave functions are 
much more complicated (see \Sec{3b}).  
Consequently, the calculation of the potential matrix elements \eq{x8} 
raises important numerical difficulties. 
At present, converged calculations are mainly restricted to elastic scattering 
\cite{MHO04,RAG08,MEO06}. 
Most breakup calculations still involve limited bases and/or simplifying assumptions 
\cite{EMO09,RAG09} but these limitations can be overcome \cite{MKY10}.

\section{Breakup reactions of two-body projectiles}\label{2b}
\setcounter{equation}{0}
\subsection{Two-cluster model}
Most of the reaction models have been applied assuming a 
two-cluster structure of the projectile. 
In this section, we specify the expression 
of the internal Hamiltonian of the projectile and 
the set of coordinates usually considered in practical applications. 
We then illustrate the models presented in \Sec{framework} and the 
approximations explained in Secs.~\ref{tdse}--\ref{cdcc} 
with various applications to the study of exotic nuclei and nuclear 
astrophysics. 

We consider here projectiles made up of a single 
fragment $f$ of mass $m_f$ and charge $Z_fe$, 
initially bound to a core $c$ of mass $m_c$ and charge $Z_ce$. 
The core and fragment are assumed to have spins $s_c$ and $s_f$. 
The internal structure of these clusters and of the target is usually neglected 
although some structure effects can be simulated by the effective potentials.

Let us now particularize the general formalism \eq{e2}--\eq{e2b} 
to the present case. 
The internal coordinates $\xi$ represent the relative coordinate 
$\vec{r} = \vec{r}_f - \vec{r}_c$. 
The structure of the projectile is 
described by the two-body internal Hamiltonian
\beq
H_0=\frac{p^2}{2\mu_{cf}}+V_{cf}(\vec{r}),
\eeqn{e201}
where $\mu_{cf}=m_cm_f/m_P$ is the reduced mass of the core-fragment pair 
(with $m_P=m_c+m_f$), $\vec{p}$ is the momentum operator of the relative motion 
and $V_{cf}$ is the potential describing the core-fragment interaction.
This potential usually includes a central part and a spin-orbit coupling term
in addition to a Coulomb potential. 
In many cases, the potential is deep enough to contain unphysical bound states 
below the ground state. 
These unphysical or forbidden states are useful because they allow 
the wave function representing the physical ground state 
to exhibit the number of nodes expected from 
the Pauli principle, as obtained in microscopic descriptions \cite{BFW77}. 
Although these forbidden states do not play any role in the core-fragment scattering, 
they could affect breakup properties. 
However, as shown in \Ref{CBM03b}, 
their presence can be ignored because their effect is negligible. 

Let $\vec{k}$ be the wave vector describing the asymptotic relative motion 
between the fragments in the projectile continuum. 
The corresponding energy is thus $E = \hbar^2 k^2/ 2 \mu_{cf}$. 
Notation $\tau$ in \Eq{e2} corresponds here to the coupling mode, 
i.e.\ to the total spin $S$ of the projectile and 
the relative orbital momentum $l$. 
The wave functions defined in \Eq{e2} read 
\beq
\phi^{JM}_{lS}(E,\vec{r}) = r^{-1} i^l [Y_l (\Omega) \otimes \chi_S]^{JM} 
u^J_{lS} (k,r),
\eeqn{e202}
where $\chi_S$ is a spinor resulting from the coupling of $s_c$ and $s_f$.
The radial functions $u^J_{lS} (k,r)$ are normalized according to 
$\la u^J_{lS} (k,r) | u^J_{lS} (k',r) \ra = \delta(k-k')$. 
The notation $\hat{k}_\xi$ in \Eq{e2a} represents the direction $\Omega_k$ of $\vec{k}$ 
and the spin orientations $\nu_c$ and $\nu_f$ of the core and fragment 
spins $s_c$ and $s_f$. 
Relation \eq{e2b} between continuum eigenstates of $H_0$ becomes 
\beq
\!\!\!\!\!\phi^{(-)}_{\Omega_k,\nu_c \nu_f}(E,\vec{r}) = 
\frac{1}{k}\sum_{lSJM} 
(s_c s_f \nu_c \nu_f | S \nu) (l S\, M\!-\!\nu\, \nu | J M) 
Y_l^{M-\nu\,*} (\Omega_k) \phi^{JM}_{lS}(E,\vec{r})
\eeqn{e203}
with the property 
$\la \phi^{(-)}_{\Omega_k,\nu_c \nu_f}(E,\vec{r}) | \phi^{(-)}_{\Omega'_k,\nu_c' \nu_f'}(E',\vec{r}) \ra 
= \delta(\vec{k}-\vec{k}') \delta_{\nu_c \nu_c'} \delta_{\nu_f \nu_f'}$. 
Notice that notations $\tau$ and $\hat{k}_\xi$ are model dependent 
and would be quite different if a tensor interaction were included in $V_{cf}$. 
A detailed description of the simple case $s_c = s_f = 0$ can be found in Ref.~\cite{Ba08}. 

\begin{figure}
\center
\begin{minipage}[b]{5.5cm}
\includegraphics[width=5cm]{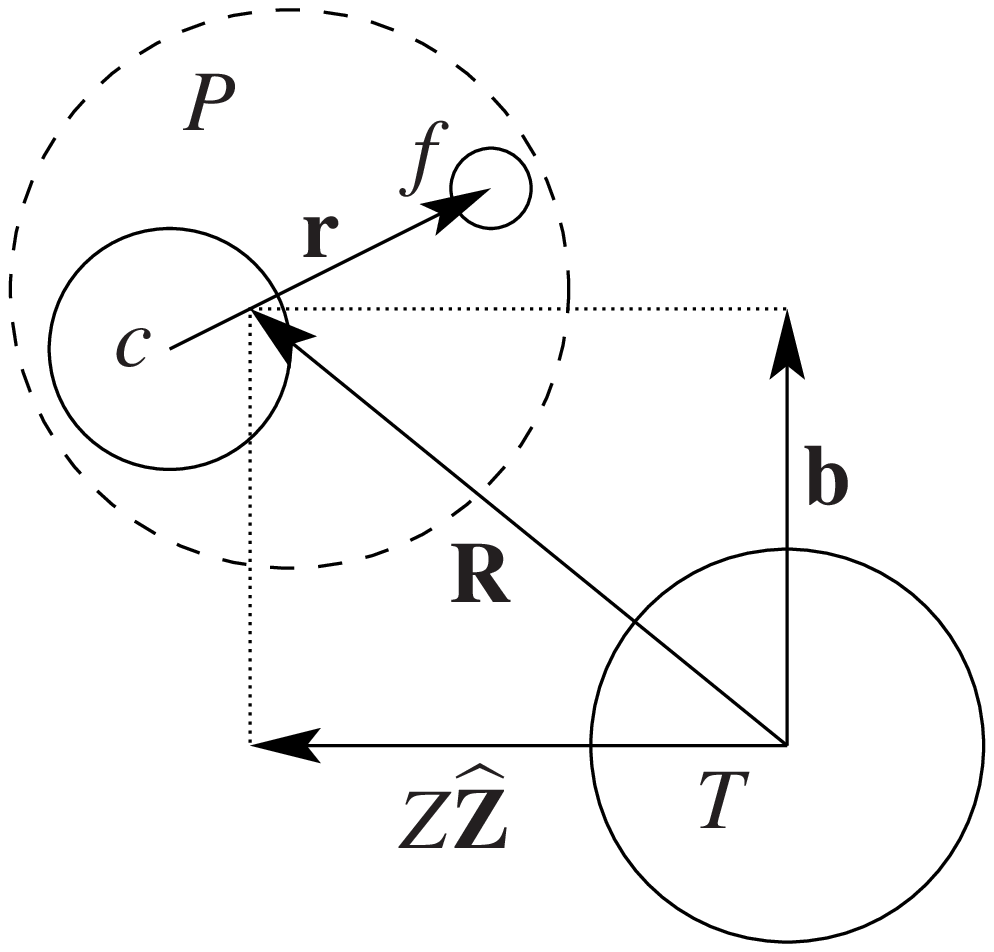}
\end{minipage} 
\begin{minipage}[b]{5.5cm}
\caption{Jacobi set of coordinates:\newline
$\vec{r}$ is the projectile internal coordinate, and\newline
$\vec{R}=\vec{b}+Z\vec{\widehat Z}$ is the target-projectile coordinate.}
\label{f20}
\end{minipage}
\end{figure}

Within this framework the description of the reaction reduces to the
resolution of a three-body \Sch equation \eq{e4} that reads,
in the Jacobi set of coordinates illustrated in \fig{f20},
\beq
\left[ \frac{P^2}{2\mu} + H_0 + V_{PT}(\vec{r},\vec{R}) \right] \Psi(\vec{r},\vec{R})
= E_{\rm T} \Psi(\vec{r},\vec{R}).
\eeqn{e204}
The projectile-target interaction \eq{e4a} then reads 
\beq
V_{PT}(\vec{R},\vec{r})&=&
V_{cT}\left(\vec{R}-\frac{m_f}{m_P}\vec{r}\right)
+V_{fT}\left(\vec{R}+\frac{m_c}{m_P}\vec{r}\right),
\eeqn{e204a}
where $V_{cT}$ and $V_{fT}$ are optical potentials that
simulate the core-target and fragment-target interactions, respectively.

For a two-body projectile, the DEA breakup cross section \eq{s16}
becomes Eq.~(46) of Ref.~\cite{GBC06}.
Integration over $\Omega_k$ and summation over $\nu_c$ and $\nu_f$
lead to the energy and angular distribution of the fragments
in the $P$-$T$ c.m.\ rest frame.
With the normalization of the positive-energy states given above,
it reads \cite{GBC06}
\beq
\frac{d\sigma_{\rm bu}}{dE d\Omega} 
= \frac{K K'}{2J_0+1} \sum_{M_0} \sum_{lJM} 
\left| \int_0^\infty b db J_{|M-M_0|}(qb) S^{(M_0)}_{klJM}(b) \right|^2, 
\eeqn{e300}
where $S^{(M_0)}_{klJM}$ are coefficients of a partial-wave expansion of
the breakup amplitude \eq{s14} (see Eq.~(43) of \Ref{GBC06}).
Breakup cross sections are mainly expressed as energy distributions $d\sigma_{\rm bu}/dE$ 
as a function of the energy of the relative motion between the fragments. 
They are obtained by integrating \eq{e300} over $\Omega$. 
However, most experimental data concern angular distributions or 
distributions of the core momentum in the laboratory frame.
Note that, in addition, theoretical results should be 
convoluted with the experimental acceptance and resolution. 
A change of frame for the theoretical results is thus in general 
not sufficient to allow a fruitful comparison with experiment. 

\subsection{Two-body breakup of exotic nuclei}
\label{tbb}
\begin{figure}
\center
\includegraphics[width=7.5cm]{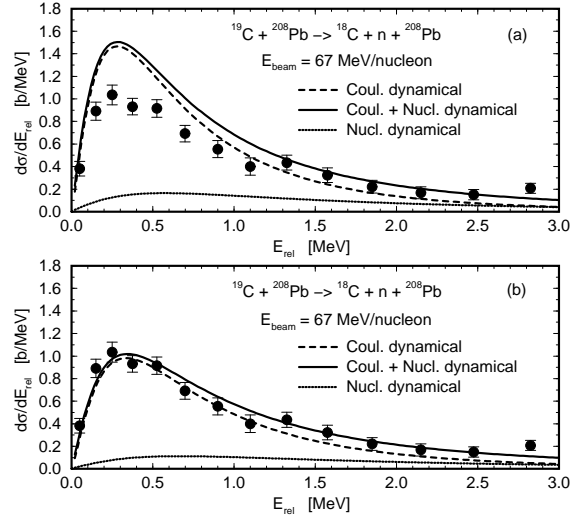}
\caption{Breakup of \ex{19}C on Pb at 67~MeV/nucleon: 
semi-classical cross sections for two different binding energies of the projectile: 
0.53 MeV (upper panel) and 0.65 MeV (lower panel) \cite{TS01r}. 
Experimental data from \Ref{NFK99}.
Reprinted figure with permission from \Ref{TS01r}.
Copyright (2001) by the American Physical Society.}
\label{f4}
\end{figure}

A first information that one can extract from experiment concerns 
the separation energy of the halo neutrons. 
Indeed, the shape of the breakup cross section and, in particular, its 
maximum are sensitive to this energy as can be shown at first 
order of perturbation theory with rather simple models based 
on the asymptotic behaviour of the halo wave function \cite{TB05}. 
An example is given by the breakup of \ex{19}C on lead at 67 MeV/nucleon. 
In \fig{f4}, a semi-classical calculation with a \ex{18}C + n two-body model shows 
that the shape of the experimental data is much better reproduced 
if the binding energy of \ex{19}C is raised from the recommended value 0.53 MeV 
to 0.65 MeV \cite{TS01r},

Indirect information can also be obtained on the spin of the ground state 
of the halo nucleus when few rather different orbital momenta are probable. 
The magnitude of the cross section is very sensitive to the orbital momentum $l$ 
of the ground state. 
A study of the one-neutron removal cross section from \ex{31}Ne 
described in a simple \ex{30}Ne + n model allows 
to rule out the prediction $7/2^-$ of the naive shell model 
and to confirm the value $3/2^-$ resulting from a shell inversion 
\cite{NKK09,HSC10}. 

\begin{figure}[t]
\center
\includegraphics[width=8.5cm]{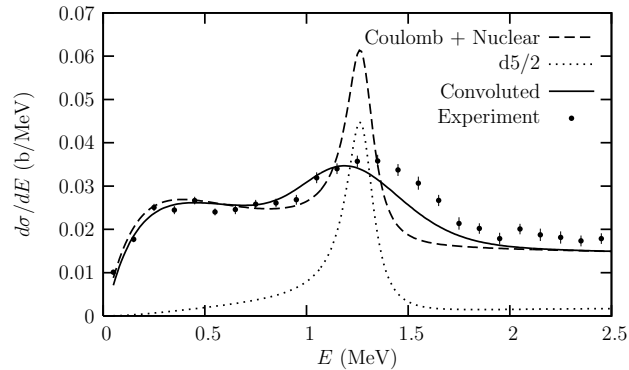}
\caption{Breakup of \ex{11}Be on a C target at 67~MeV/nucleon: 
calculation performed in a semi-classical model \cite{CGB04}.
Experimental data from \Ref{FNA04}.}
\label{f6}
\end{figure}

Nuclear-induced two-body breakup on light targets is an interesting tool 
to observe resonances of a halo nucleus and to assess some of their properties. 
In \fig{f6} are displayed experimental data on the \ex{11}Be breakup on a C target 
at 67~MeV/nucleon \cite{FNA04}. 
These data present a broad bump near the location of a known resonance 
with an assumed spin-parity $5/2^+$. 
The bump width is however broader than the known resonance width. 
A semi-classical calculation (dashed line) based on a \ex{10}Be + n model
reproduces the shape of the data very well 
after convolution with the experimental resolution (full line). 
Moreover the $d5/2$ component of the theoretical cross section 
(dotted line) resonates and confirms the $5/2^+$ attribution. 

Breakup reactions are also used to infer the spectroscopic factor
of the dominant configuration in the core+nucleon structure
of halo nuclei \cite{Nak94,HT03}.
Various theoretical studies have been performed to assess the sensitivity
of breakup calculations to the projectile description \cite{CN06,CN07}.
These studies have revealed that the breakup cross sections not only depend
on the initial bound state of the projectile, but are also sensitive to
the description of its continuum \cite{CN06}. 
Moreover it has been shown that, for loosely-bound projectiles, 
only the tail of the wave function is probed in the breakup process 
and not its whole range \cite{CN07,GBB06}. 
These studies therefore suggest that one should proceed with caution 
when extracting spectroscopic factors of weakly-bound nuclei 
from breakup measurements, as other structure properties, 
like the continuum description, may hinder that extraction. 

\begin{figure}[t]
\center
\includegraphics[width=8.5cm]{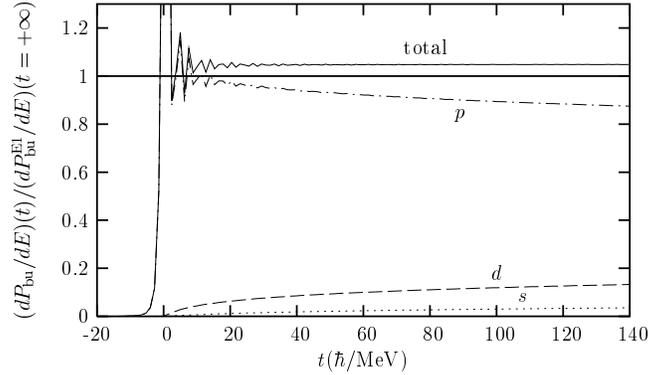}
\caption{Influence of the couplings inside the continuum \cite{CB05}.
Time evolution of the numerical breakup probability per energy unit \eq{e8a}
for \ex{11}Be impinging on Pb at about 45~MeV/nucleon 
for a \ex{10}Be-n relative energy $E=1.5$ MeV and an
impact parameter $b=100$ fm.
Reprinted figure with permission from \Ref{CB05}.
Copyright (2005) by the American Physical Society.}
\label{f5}
\end{figure}

As mentioned earlier, many Coulomb-breakup experiments
have been analyzed within the framework of the first order
of the perturbation theory (see \Sec{FO}).
In order to assess the validity of that approximation,
various authors have compared perturbation calculations 
to numerical resolutions of the time-dependent \Sch
equation \cite{EB96,TB01,EBS05,CB05}.
These studies have shown that, in many cases, breakup
cannot be modelled as a one-step process from the initial bound state
towards the continuum and that higher-order effects should
be considered for a reliable description of the reaction.
In particular, they indicate that significant couplings 
are at play inside the continuum.
To illustrate this, \fig{f5} displays the time evolution of 
the breakup probability per energy unit \eq{e8a} for the collision of \ex{11}Be 
on Pb at about 45~MeV/nucleon computed within the time-dependent \ex{10}Be + n  
model of \Ref{CBM03c}. 
The obtained value is divided by 
its evaluation at the first-order of the perturbation theory \eq{e11c} 
at $t\rightarrow+\infty$. 
After a sharp increase at the time of closest approach $t=0$,
the breakup probability (full line) oscillates and then stabilizes at a value 
which differs by about 5\% from its first-order estimate. 
Although the total breakup probability becomes stable,
its partial-wave composition still varies: the dominant $p$ wave contribution 
(dash-dotted line) is depleted towards the $s$ (dotted line) 
and especially $d$ (dashed line) ones.
This signals couplings inside the continuum, which may
affect the evaluation of breakup observables \cite{CB05,EBS05}.
We will see in the next section that it may perturb the analysis
of breakup reactions of astrophysical interest \cite{Oga06,GCB07,SN08,Esb09}.

\subsection{Application to nuclear astrophysics}
Radiative-capture reactions are a crucial ingredient in the determination 
of the reaction rates in nuclear astrophysics. 
However the difficulty of their measurement and, in some cases, the scatter of the results 
has raised interest in indirect methods where the time-reversed reaction 
is simulated by virtual photons in the Coulomb field of a heavy nucleus \cite{BBR86,BR96}. 
The radiative-capture cross section can be extracted from breakup cross sections 
if one assumes that the dissociation is due to E1 virtual photons and occurs in a single step. 
A typical example is the \ex{7}Be(p,$\gamma$)\ex{8}B reaction which has been studied 
with the breakup of $^8$B into $^7$Be+p on heavy targets at different energies 
\cite{Mot94,Kik97,Dav98,Gui00,KGP01,Dav01l,Dav01c,Sch06}. 

Though appealing, the breakup method also faces a number of difficulties. 
First, while many reactions are dominated by an E1 transition, 
an E2 contribution to the breakup cross section may not be negligible \cite{EB96}. 
Second, higher-order effects, i.e.\ transitions from the initial bound 
state into the continuum through several steps may not be negligible \cite{EB96,TB01,EBS05,CB05}. 
Finally, the nuclear interactions between the projectile and the target may interfere 
with the Coulomb interaction \cite{TS01r,CBM03c}. 
Therefore elaborate reaction theories must be used to interpret the experimental data. 

\begin{figure}
\center
\includegraphics[width=10cm]{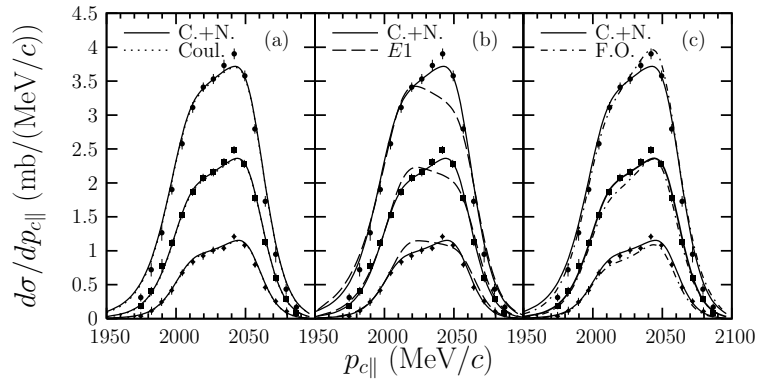}
\caption{\ex{8}B Coulomb breakup on Pb at 44~MeV/nucleon.
Parallel-momentum distribution of the \ex{7}Be core corresponding
to various angular cuts calculated in a DEA model \cite{GCB07}.
(a) Influence of nuclear and Coulomb interactions on the calculation.
(b) Effects of the various multipoles of the Coulomb interaction.
(c) Role of the higher-order effects on the calculation.
Experimental data from \Ref{Dav98}.}
\label{f8}
\end{figure}

The experiments on the breakup of \ex{8}B have been analyzed in a number of papers 
\cite{TNT01,MTT02,DT03,SN05,Oga06,GCB07}.
Figure \ref{f8} shows a comparison between the experimental data of \Ref{Dav98} 
and DEA calculations \cite{GCB07}.
Without adjustable parameters, the calculations (full lines) fairly reproduce the asymmetry 
exhibited by the data which could not be well explained in earlier works \cite{MTT02,SN05}. 
The three panels of \fig{f8} illustrate the influence of various
approximations upon the calculation \cite{GCB07}.
The left panel illustrates that nuclear $P$-$T$ interactions can
be neglected when data are restricted to forward angles.
The central panel confronts a dynamical calculation including
only the dipole term of the Coulomb interaction (dashed lines)
to the full calculation, indicating that higher multipoles
have a significant effect on the breakup process.
The right panel compares the dynamical calculation to
its first-order approximation (dot-dashed lines), emphasizing the necessity
to include higher-order effects in breakup calculations.
These results show that some of the assumptions of
the breakup method \cite{BBR86,BR96} are not valid.
It is therefore difficult to infer the accuracy of the $S$ factors
extracted from breakup cross sections.

\begin{figure}[t]
\center
\includegraphics[width=8cm]{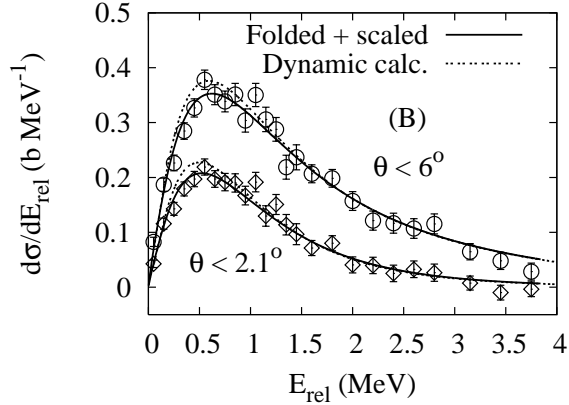}
\caption{Breakup of \ex{15}C on Pb at 68~MeV/nucleon.
The experimental energy distribution 
measured for two scattering-angle cuts \cite{Nak09} is confronted to the
time-dependent calculation of \Ref{Esb09}.
Reprinted figure with permission from \Ref{Esb09}.
Copyright (2009) by the American Physical Society.}
\label{f7}
\end{figure}

An interesting problem was raised by the \ex{14}C(n,$\gamma$)\ex{15}C capture reaction. 
The measured cross sections for the Coulomb breakup of \ex{15}C \cite{Hor02,Dat03}
provided an $S$ factor which disagreed with direct measurements \cite{Rei05,Rei08}. 
Moreover, theoretical analyses indicated that the Coulomb-breakup cross sections were inconsistent 
with information obtained from \ex{15}F by charge symmetry and with 
microscopic models \cite{TBD06}. 
A new measurement \cite{Nak03,Nak09} has obtained breakup cross sections that fully agree 
with properties of the mirror system and with theory \cite{SN08,Esb09}.
These theoretical analyses show that a fully dynamical calculation,
taking proper account of higher-order effects
is necessary to correctly analyze the breakup measurements,
in agreement with the analysis of the \ex{8}B Coulomb breakup of \Ref{GCB07}. 
They also indicate that including both Coulomb and nuclear
interactions as well as their interferences is necessary to correctly
reproduce data at large scattering angles.
In this way a very good agreement can be obtained between
direct and indirect measurements of the $S$ factor.
\fig{f7} displays the breakup cross section of \ex{15}C on Pb
measured at 68~MeV/nucleon \cite{Nak09}
and its comparison to the theoretical calculation of the
time-dependent model of \Ref{Esb09}.
The dotted lines show the direct results of the calculation,
while the full lines correspond to these results folded by the
experimental resolution and scaled to the data.

\section{Breakup reactions of three-body projectiles}
\label{3b}
\setcounter{equation}{0}
\subsection{Three-cluster model of projectile}
Let us consider a system of three particles, 
the core with coordinate $\vec{r}_c$, mass $m_c$ and charge $Z_c e$ 
and two fragments with coordinates $\vec{r}_1$ and $\vec{r}_2$, 
masses $m_1$ and $m_2$, and charges $Z_1 e$ and $Z_2 e$. 
The projectile mass is $m_p = m_c + m_{12}$ with $m_{12} = m_1 + m_2$. 
After removal of the c.m.\ kinetic energy $T_{\rm c.m.}$, 
the three-body Hamiltonian of this system in Eq.~\eq{e2} can be written as 
\beq
H_0 = \frac{p_c^2}{2m_c} + \frac{p_1^2}{2m_1} + \frac{p_2^2}{2m_2} 
+ V_{c1} + V_{c2} + V_{12} - T_{\rm c.m.},
\eeqn{z0}
where $V_{ij}$ is an effective potential between particles $i$ and $j$ ($i,j = c,1,2$). 
We assume that these interactions involve central, spin-orbit and Coulomb terms. 
These potentials may contain unphysical bound states below the two-cluster ground state 
to simulate effects of the Pauli principle. 
These forbidden states must be eliminated from the three-body wave functions 
either with pseudopotentials \cite{KP78} or with supersymmetric transformations 
\cite{Ba87,Ba87b}. 

Various resolution techniques can be considered for obtaining the 
wave functions of a three-body projectile. 
A first option is to describe this projectile with an expansion in 
Gaussian functions depending on Jacobi coordinates \cite{HKK03,SLY03,EMO09}. 
For bound states, the wave functions can be obtained from a variational calculation. 
Well established techniques allow systematic calculations of the matrix elements 
\cite{HKK03,SLY03}. 
Calculations are then simpler when the interactions are expressed in terms of Gaussians. 
At negative energies, this type of expansion may however have convergence problems 
in the description of extended halos. 
At positive energies, it is convenient to obtain pseudostates but not convenient 
to obtain scattering states. 

Let us describe another efficient tool to deal with three-body systems, 
the formalism of hyperspherical coordinates. 
It is especially interesting when the two-cluster subsystems are unbound 
so that only a three-body continuum exists.  
Notation $\xi$ of \Sec{framework} represents here 
five angular variables and one coordinate with the dimension of a length, 
the hyperradius (see Refs.~\cite{RR70,ZDF93,DDB03} for details). 
Four angular variables correspond to physical angles and the fifth one is 
related to a ratio of coordinates defined below in Eq.~\eq{z5}. 
The wave functions are expanded in series of hyperspherical harmonics, 
i.e.\ a well known complete set of orthonormal functions of the five angular variables. 
The coefficients are functions of the hyperradius and can be obtained from 
variational calculations. 
Scattering states can be obtained from extensions of the $R$ matrix theory 
\cite{DTB06,DB10}.
A drawback of this method is that the hyperspherical expansion may converge rather slowly. 

With the dimensionless reduced masses $\mu_{c(12)} = m_c m_{12}/m_P m_N$ 
and $\mu_{12} = m_1 m_2$ $/m_{12}m_N$ where $m_N$ is the nucleon mass for example, 
the internal coordinates $\xi$ are scaled Jacobi coordinates defined as 
\beq
\vec{x} = \sqrt{\mu_{12}} (\vec{r}_2 - \vec{r}_1)
\eeqn{z1}
and 
\beq
\vec{y} = \sqrt{\mu_{c(12)}} \left( \vec{r}_c - \frac{m_1 \vec{r}_1 + m_2 \vec{r}_2}{m_{12}} \right),
\eeqn{z2}
i.e., up to a scaling factor, the relative coordinate between the clusters 1 and 2  
and the relative coordinate of their centre of mass with respect to the core. 
With Laplacians $\Delta_x$ and $\Delta_y$ with respect to $\vec{x}$ and $\vec{y}$,
the Hamiltonian \eq{z0} of this three-body projectile can be rewritten as 
\beq
H_0 = -\frac{\hbar^2}{2m_N} (\Delta_x + \Delta_y) + V_{c1} + V_{c2} + V_{12}.
\eeqn{z3}
To investigate the breakup cross sections for this system, 
we need wave functions at both positive and negative energies. 

In the notation of Refs.~\cite{RR70,DDB03}, 
the hyperradius $\rho$ and hyperangle $\alpha$ are defined as 
\beq
\rho = \sqrt{x^2+y^2}
\eeqn{z4}
and 
\beq
\alpha = \arctan (y/x).
\eeqn{z5} 
The hyperangle $\alpha$ and the orientations $\Omega_x$ and $\Omega_y$ of $\vec{x}$ and $\vec{y}$ 
provide a set of five angles collectively denoted as $\Omega_5$. 
The volume element is $d\vec{x} d\vec{y} = \rho^5 d\rho d\Omega_5$ with  
$d\Omega_5 = \sin^2 \alpha \cos^2 \alpha d\alpha d\Omega_x d\Omega_y$. 

The hyperspherical harmonics form an orthonormal basis which verifies a closure relation. 
The purely spatial hyperspherical harmonics read \cite{RR70,DDB03} 
\beq
{\cal Y}^{l_x l_y}_{KLM_L}(\Omega_5)=\phi^{l_x l_y}_{K}(\alpha) 
\left[ Y_{l_x}(\Omega_x)\otimes Y_{l_y}(\Omega_y) \right] ^{LM_L}.
\eeqn{z6b}
where $K$ is the hypermomentum quantum number, $l_x$ and $l_y$ are the orbital quantum numbers
associated with $\vec{x}$ and $\vec{y}$, and $L$ is the quantum number of total orbital momentum.
The functions $\phi^{l_x l_y}_{K}$ depending on the hyperangle $\alpha$ are 
defined in Eqs.~(9) and (10) of Ref.~\cite{DDB03}. 
The hyperspherical harmonics involving spin are defined by 
\beq
{\cal Y}^{JM}_{\gamma K}(\Omega_5)=\left[ {\cal Y}^{l_x l_y}_{KL}(\Omega_5) \otimes \chi_{S}
\right] ^{JM},
\eeqn{z6a}
where $\chi_S$ is a spinor corresponding to a total spin $S$ of the three clusters. 
Intermediate couplings as, for example, the total spin $s_{12}$ of the fragments 
are not displayed for simplicity. 
Index $\gamma$ stands for $(l_x l_y L S)$. 

A partial wave function $\phi^{JM}$ is a solution of the \Sch equation \eq{e2} 
associated with the three-body Hamiltonian \eq{z3} at energy $E$. 
It can be expanded as 
\beq
\phi^{JM}(E,\rho,\Omega_5) = \rho^{-5/2} \sum_{\gamma K} {\chi}^{J}_{\gamma K}(E,\rho)\ 
{\cal Y}^{JM}_{\gamma K}(\Omega_5),
\eeqn{z6}
For bound states ($E<0$), the hyperradial wave functions decrease asymptotically as 
\beq
{\chi}^{J}_{\gamma K}(E,\rho) \mathop{\sim} \limits_{\rho \rightarrow \infty} 
\exp (-\sqrt{2m_N |E|/\hbar^2} \rho).
\eeqn{z9}
Index $\tau$ of \Eq{e2} is irrelevant for bound states within the present assumptions. 
The normalization of the scattering states ($E>0$) is fixed by their asymptotic form. 
Several choices are possible. 
The asymptotic form of the hyperradial scattering wave function is for instance 
given by \cite{BCD09}
\beq
\lefteqn{{\chi}^{J}_{\gamma K (\gamma_\omega K_\omega)}(E,\rho) 
\mathop{\rightarrow} \limits_{\rho \rightarrow \infty} 
i^{K_\omega+1}(2\pi/k)^{5/2} }\nonumber\\
&\times&
\left[ H^-_{K+2} (k\rho) \delta_{\gamma \gamma_\omega}\delta_{KK_\omega}
-U^{J}_{\gamma K,\gamma_\omega K_\omega} H^+_{K+2} (k\rho) \right],
\eeqn{z11}
where $k=\sqrt{2m_N E/\hbar^2}$ is the wave number 
and $H^{-}_{K}$ and $H^{+}_{K}$ are incoming and outgoing functions 
\cite{DTV98,TDE00,DTB06}. 
In the neutral case, i.e.\ when clusters 1 and 2 are neutrons, these functions read 
$H^{\pm}_{K} (x)=\pm i ( \pi x/2 )^{1/2}
\left[ J_{K}(x)\pm i Y_{K}(x) \right]$ 
where $J_K$ and $Y_K$ are Bessel functions of first and second kind, respectively.
In the charged case, expression \eq{z11} is only an approximation because the asymptotic form
of the Coulomb interaction is not diagonal in hyperspherical coordinates \cite{VNA01,DTB06}.
The indices $\gamma_\omega K_\omega$ denote the partial entrance channels for this solution. 
For scattering states, index $\tau$ of \Eq{e2} is necessary and rather complicated: 
it represents the entrance channel $\gamma_\omega K_\omega$. 
The asymptotic behaviour of a given partial wave depends on the collision matrix. 
For real interactions, the collision matrix $\vec{U}^{J}$ of each partial wave $J$ is unitary and symmetric. 
For three-body scattering, it differs from two-body collision matrices in an important aspect: 
its dimension is infinite since the particles can share the angular momentum in an infinity of ways. 
In practical calculations, its dimension depends on the number of hypermomenta included 
in the calculation, limited to a maximum $K$ value, denoted as $K_{\rm max}$. 

The three-body final scattering states are described asymptotically with two relative 
wave vectors. 
Let $\vec{k}_c$, $\vec{k}_1$, $\vec{k}_2$ be the wave vectors of the core and fragments
in the projectile rest frame. 
The asymptotic relative motions are defined by the relative wave vector of the neutrons 
\beq
\vec{k}_{21} = \sqrt{\mu_{12}} \vec{k}_x = \frac{m_1 \vec{k}_2 - m_2 \vec{k}_1}{m_{12}}  
\eeqn{z12}
and the relative wave vector of the core with respect to the centre of mass 
of the fragments 
\beq
\vec{k}_{c(12)} = \sqrt{\mu_{c(12)}} \vec{k}_y = 
\frac{m_{12} \vec{k}_c - m_c (\vec{k}_1 + \vec{k}_2)}{m_P}.
\eeqn{z13}
The total internal energy of the projectile with respect to the 
three-particle threshold is given by 
\beq
E = \frac{\hbar^2}{2m_N} k^2 = \frac{\hbar^2}{2m_N} (k_x^2 + k_y^2).
\eeqn{z10}
The orientations $\Omega_{k_x}$ of $\vec{k}_{x}$ and $\Omega_{k_y}$ of $\vec{k}_{y}$ 
and the ratio $\alpha_k = \arctan(y_k/x_k)$ form the wave vector hyperangles $\Omega_{5k}$. 
The hyperangle $\alpha_k$ controls the way the projectile energy $E$ is shared among the fragments. 
For example, the energy of the relative motion between fragments 1 and 2 is $E \cos^2 \alpha_k$.
In the scattering states \eq{e2a}, notation $\hat{k}_\xi$ thus represents $\Omega_{5k}$ 
and the final orientations $\nu_c$, $\nu_1$, $\nu_2$ of the three spins. 
It is convenient to replace these orientations by the total spin $s_{12}$ 
of the fragments, the total spin $S$ and its projection $\nu$. 
Relation \eq{e2b} is then given by 
\beq 
\phi^{(-)}_{\Omega_{5k} S\nu} (E, \rho, \Omega_5)
= (2\pi)^{-3} \rho^{-5/2} \sum_{JM} \sum_{l_{x\omega} l_{y\omega} L_\omega K_\omega} 
(L_\omega S\, M\!-\!\nu\, \nu|JM) 
\eol \times
{\cal Y}_{l_{x\omega} l_{y\omega} K_\omega}^{L_\omega M-\nu*} (\Omega_{5k}) 
\phi^{JM}_{\gamma K (\gamma_\omega K_\omega)}(E,\rho,\Omega_5).
\eeqn{z14}
where $\gamma_\omega = (l_{x\omega}, l_{y\omega}, L_\omega, S)$. 
These functions are normalized with respect to 
$\delta (\vec{k_x}-\vec{k'_x}) \delta (\vec{k_y}-\vec{k'_y}) \delta_{SS'} \delta_{\nu\nu'}$.

The hyperradial wave functions ${\chi}^{J}_{\gamma K}$ are to be determined 
from the \Sch equation \eq{e2}. 
The parity $\pi=(-1)^K$ of the three-body relative motion restricts the sum over $K$ 
to even or odd values. 
Rigorously, the summation over $\gamma K$ in \eq{z6} should contain an infinite number of terms. 
In practice, this expansion is limited by 
the truncation value $K_{\rm max}$.
The $l_x$ and $l_y$ values are limited by $l_x + l_y \le K \le K_{\rm max}$. 
For weakly-bound and scattering states, it is well known that the convergence is rather slow 
and that large $K_{\rm max}$ values must be used. 

The functions ${\chi}^{J}_{\gamma K}$ are derived from a set of coupled differential equations 
\cite{DDB03,DTB06} 
\beq
\left[-\frac{\hbar^2}{2m_N} \left( \frac{d^2}{d\rho^2} - \frac{(K+3/2)(K+5/2)}{\rho^2}\right) -E \right]
{\chi}^{J}_{\gamma K}(E,\rho) &&
\eol
+ \sum_{\gamma' K'} V^{J}_{\gamma' K',\gamma K}(\rho)\, {\chi}^{J}_{\gamma' K'}(E,\rho)=0, &&
\eeqn{z7} 
where the potentials matrix elements are defined as
\beq
V^{J}_{\gamma' K',\gamma K}(\rho)=
\langle {\cal Y}^{JM}_{\gamma' K'}(\Omega_5) | \sum_{i>j=1}^3 V_{ij}(\vec{r}_j-\vec{r}_i) |
{\cal Y}^{JM}_{\gamma K}(\Omega_5) \rangle.
\eeqn{z8} 
For bound states, approximate solutions 
can be obtained with an expansion on a finite square-integrable basis. 
However, using such a basis for scattering states raises problems 
since they do not vanish at infinity. 
Their asymptotic form requires a proper treatment. 
This technical difficulty can be solved within the $R$-matrix theory \cite{LT58,DTB06,DB10}  
which allows matching a variational function over a finite interval 
with the correct asymptotic solutions of the \Sch equation. 

In the $R$-matrix approach, both bound and scattering hyperradial wave functions are approximated 
over the internal region by an expansion on a set of square-integrable variational functions defined over $[0,a]$. 
Lagrange-mesh basis functions are quite efficient for describing two-body bound and scattering states 
\cite{BH86,HSV98,HRB02}. 
The main advantage of this technique is to strongly simplify the calculation of matrix elements \eq{z8} 
without loss of accuracy if the Gauss approximation consistent with the mesh is used \cite{DDB03}. 
This method was extended to three-body bound states in Ref.~\cite{DDB03} 
and to three-body scattering states in Ref.~\cite{DTB06}. 
We refer the reader to those references for details. 

\subsection{Dipole strength distribution}
\label{E1s}
The E1 strength distribution for transitions from the ground state to the continuum 
is a property of the projectile that can be extracted from breakup experiments 
under some simplifying assumptions for cases where E1 is dominant. 
In the hyperspherical coordinate system, the multipole operators are given by Eq.~(B2) 
of Ref.~\cite{DDB03}. 
For example, in two-neutron halo nuclei, the E1 strength is given by 
\beq
{\cal M}_\mu^{\rm E1} (\rho, \Omega_5) = e Z_c \frac{m_{12}}{m_P}\, 
\frac{\rho \sin \alpha}{\sqrt{\mu_{c(12)}}}\, Y_1^\mu (\Omega_y).
\eeqn{z15}
The E1 transition strength \eq{e11d} from the ground state at negative energy $E_0$ 
with total angular momentum $J_0$ to the continuum is given by 
\beq
\frac{dB({\rm E1})}{dE} 
& = & \frac{4}{2J_0+1} \left( \frac{m_{\rm N}}{\hbar^2} \right) E^2 
\sum_{M_0 \mu} \sum_{S \nu} \int d\Omega_{5k} 
\eol &&
\left| \left\langle \phi^{(-)}_{\Omega_{5k} S\nu} (E, \rho, \Omega_5) | 
{\cal M}^{{\rm E}1}_{\mu}(\rho, \Omega_5)
| \phi^{J_0M_0}(E_0, \rho, \Omega_5) \right\rangle \right|^2.
\eeqn{z16}

The E1 strength presents the advantage that it can also be calculated in various ways 
without constructing the complicated three-body scattering states \cite{HSB10}. 
Most model calculations of the E1 strength for \ex{6}He indicate 
a concentration of strength at low energies $E$ 
\cite{BCD09,RAG09,DTV98,CFJ97,MKA01,HS07}. 
The origin of this low-energy bump remains unclear and can sometimes be attributed 
to a three-body resonance \cite{DTV98,BCD09}.
The existence of such a bump does not agree with the GSI data \cite{Aum99}.

\begin{figure}
\center
\includegraphics[width=8cm]{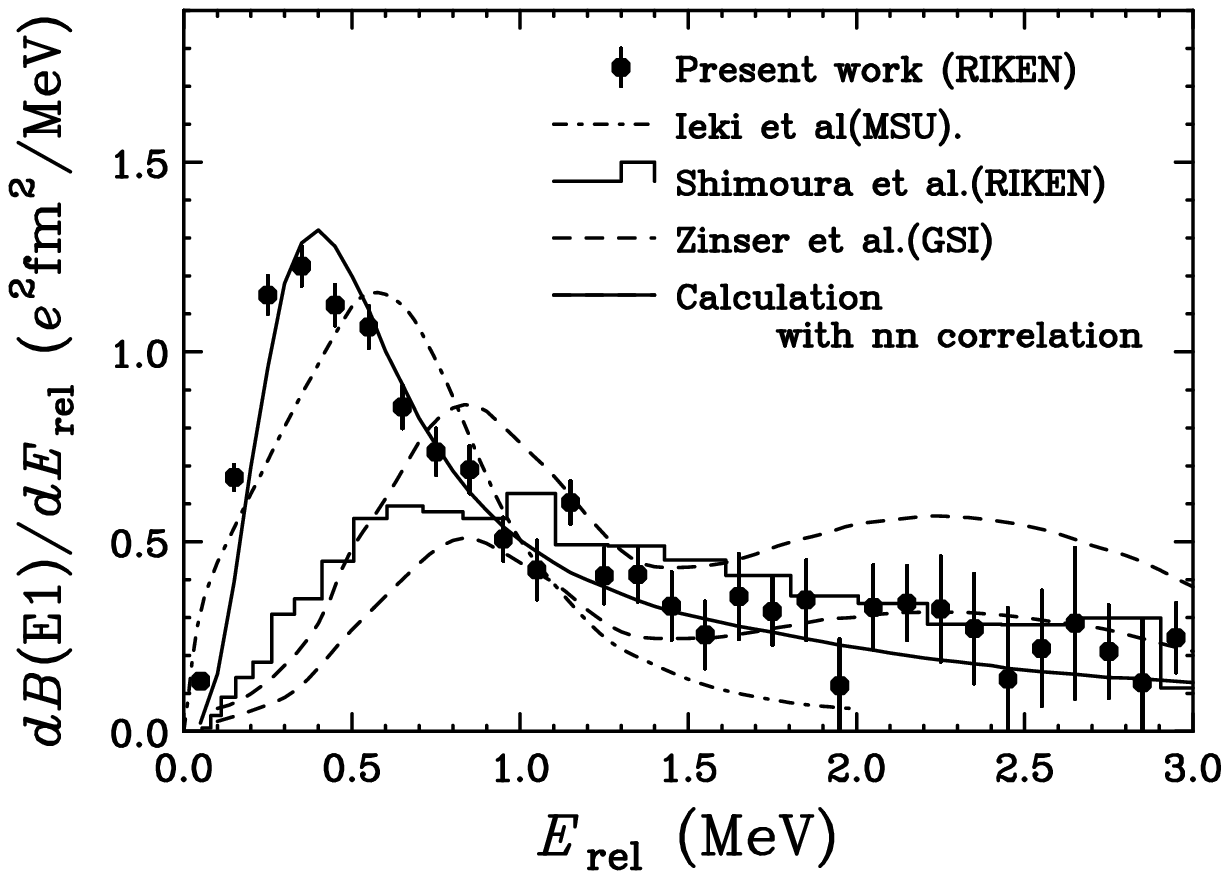}
\caption{Experimental E1 strength for the breakup of the \ex{11}Li two-neutron halo nucleus: 
\Ref{NVS06}  (full circles), \Ref{ISG93} (dash-dotted line), \Ref{SNI95} (histogram), \Ref{ZHN97} (dashed line).
Reprinted figure with permission from \Ref{NVS06}.
Copyright (2006) by the American Physical Society.}
\label{f12b}
\end{figure}

This puzzling problem deserves further studies. 
A first-order description of Coulomb breakup for \ex{6}He is probably 
not very accurate (see \Sec{ccehe6}), even at the energies of the GSI experiment \cite{Aum99}. 
Extracting the E1 strength from breakup measurement is very difficult and not without ambiguities. 
This is exemplified by the variety of experimental results obtained for the breakup 
of the \ex{11}Li two-neutron halo nucleus. 
As shown in \fig{f12b},  most early experiments \cite{ISG93,SNI95,ZHN97} did not display 
a significant strength at low energies 
in contradiction with data from the more recent RIKEN experiment \cite{NVS06}. 

\subsection{The CCE approximation for three-body projectiles}
\label{ccehe6}
We consider a collision between a three-body projectile and a structureless 
target with mass $m_T$ and charge $Z_T e$ \cite{BCD09}.  
The breakup reaction is described by the four-body \Sch equation \eq{e4} 
where $H_0$ is given by \Eq{z3}. 
The effective potential \eq{e4a} between projectile and target is defined as 
\beq
V_{PT} (\vec{R},\vec{x},\vec{y}) & = 
& V_{cT} \left( \vec{R} + \frac{m_{12}}{m_P}\,\frac{\vec{y}}{\sqrt{\mu_{c(12)}}} \right) 
+ V_{1T} \left( \vec{R} - \frac{m_c}{m_P}\, \frac{\vec{y}}{\sqrt{\mu_{c(12)}}} 
- \frac{m_2}{m_{12}}\, \frac{\vec{x}}{\sqrt{\mu_{12}}} \right) 
\eol
&& + V_{2T} \left( \vec{R} - \frac{m_c}{m_P}\, \frac{\vec{y}}{\sqrt{\mu_{c(12)}}} 
+ \frac{m_1}{m_{12}}\, \frac{\vec{x}}{\sqrt{\mu_{12}}} \right).
\eeqn{z20}
In this expression, each interaction $V_{iT}$ between a constituent of the projectile and the target 
is simulated by a complex optical potential 
(including a possible Coulomb interaction taking the cluster extension into account). 

In order to obtain breakup cross sections, one must calculate transition matrix elements 
for the breakup into three fragments. 
The transition matrix elements \eq{s7} read 
\beq
T_{fi} & = & (\mu_{12} \mu_{c(12)})^{-3/4} \eol
 & \times & \langle e^{i\vec{K}' \cdot \vec{R}} \phi^{(-)}_{\Omega_{5k} S\nu} (E, \rho, \Omega_5)
| V_{PT} (\vec{R},\vec{x},\vec{y}) | \Psi^{(M_0)}(\vec{R}, \rho, \Omega_5) \rangle
\eeqn{z21}
for four-body breakup. 
The factor $(\mu_{12} \mu_{c(12)})^{-3/4}$ appears when the integration is performed 
in coordinates $\rho$ and $\Omega_5$ and the bound-state wave function \eq{z6} is normed 
in this coordinate system rather than in Jacobi coordinates \cite{BCD09}. 
At the eikonal approximation, the exact scattering wave function $\Psi$ in \Eq{z21} 
is replaced by its approximation given by Eqs.~\eq{e20} and \eq{e26}.
The transition matrix element \eq{z21} is then obtained following \Eq{s13} as 
\beq
T_{fi} = i\hbar v \int d\vec{b} e^{-i \vec{q}.\vec{b}} S^{(M_0)}_{S\nu} (E,\Omega_{5k},\vec{b}).
\eeqn{z22a}
with the eikonal breakup amplitude \eq{s14}, that reads here \cite{BCD09}
\beq
S^{(M_0)}_{S\nu} (E,\Omega_{5k},\vec{b}) & = & (\mu_{12} \mu_{c(12)})^{-3/4} 
\eol & \times & 
\langle \phi^{(-)}_{\Omega_{5k} S\nu} (E, \rho, \Omega_5)| 
e^{i \chi(\vec{b},\vec{s}_x,\vec{s}_y)} | \phi^{J_0 M_0}(E_0,\rho, \Omega_5) \rangle.
\eeqn{z22}
Following \Eq{z20}, the eikonal phase shift $\chi$ defined in \Eq{e27} is obtained as 
\beq
\chi = \chi_{cT} + \chi_{1T} + \chi_{2T}.
\eeqn{z23}
It depends on the transverse part $\vec{b}$ of $\vec{R}$ as well as on the transverse parts 
$\vec{s}_x$ and $\vec{s}_y$ of the scaled Jacobi coordinates $\vec{x}$ and $\vec{y}$. 

From the transition matrix elements \eq{z21}, various cross sections can be derived. 
The differential cross section \eq{s16} with respect to the eight independent variables 
$\Omega$, $\vec{k}_{21}$, $\vec{k}_{c(12)}$ reads in the c.m.\ frame 
\beq
\frac{d\sigma}{d\Omega d\vec{k}_{21} d\vec{k}_{c(12)}} 
= \frac{1}{2J_0+1}\, \frac{1}{4\pi^2} \left( \frac{\mu}{\hbar^2} \right)^2 
\frac{K'}{K} \sum_{S \nu M_0} |T_{fi}|^2. 
\eeqn{z24}
The physical wave numbers $k_{21}$ and $k_{c(12)}$ are proportional 
to $k_x$ and $k_y$ and can thus be expressed from $k$ and $\alpha_k$ \cite{BCD09}. 
Integrating \Eq{z24} over all angles $\Omega$ and $\Omega_{5k}$ 
leads to the energy distribution cross section $d\sigma/dE$.

\begin{figure}
\setlength{\unitlength}{1mm}
\begin{picture}(140,50) (0,0) 
\put(20,-30){\mbox{\scalebox{1.0}{\includegraphics[width=9cm]{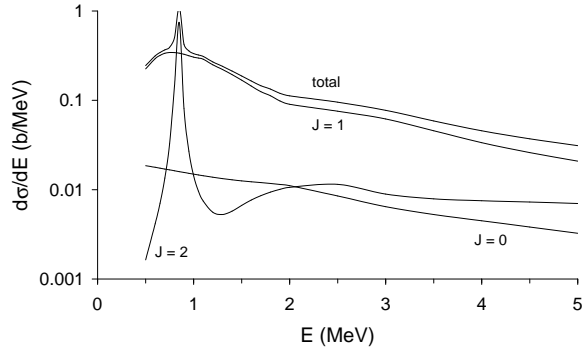}}}}
\end{picture} \\
\caption{CCE calculation of the total and $0^+$, $1^-$, $2^+$ partial cross sections 
of $^6$He breakup on $^{208}$Pb at 240 MeV/nucleon \cite{BCD09}.
Reprinted figure with permission from \Ref{BCD09}.
Copyright (2009) by the American Physical Society.}
\label{f10a}
\end{figure}

The CCE approximation has allowed calculating various elastic and breakup cross sections 
for \ex{6}He on \ex{208}Pb by treating \ex{6}He as an $\alpha$ + n + n three-body system \cite{BCD09}. 
In \fig{f10a}, the contribution from the different partial waves is displayed at 240 MeV/nucleon. 
As expected for a transition from a $0^+$ ground state, the $J = 1^-$ component is dominant. 
However the $J=0^+$ and $J=2^+$ components are not negligible. 
The known $2^+$ resonance at 0.82 MeV is clearly visible in the total cross section. 
Extracting an E1 strength from such data is thus not easy, even at this high energy. 

\begin{figure}
\setlength{\unitlength}{1mm}
\begin{picture}(140,50) (0,0) 
\put(20,-30){\mbox{\scalebox{1.0}{\includegraphics[width=9cm]{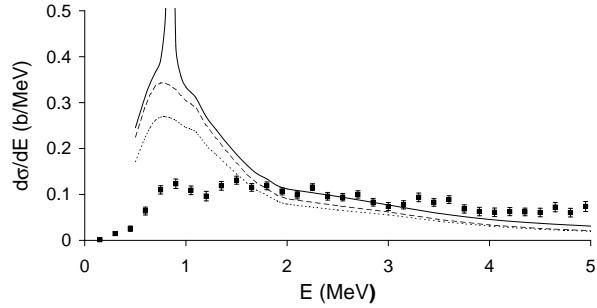}}}}
\end{picture} \\
\caption{Comparison \cite{BCD09} between the total CCE cross section (full line) 
of $^6$He breakup on $^{208}$Pb at 240 MeV/nucleon with the experimental data of Ref.~\cite{Aum99}. 
The $1^-$ partial cross sections calculated with two types of elimination of forbidden states 
(supersymmetry: dashed line, projection: dotted line) are also displayed.
Reprinted figure with permission from \Ref{BCD09}.
Copyright (2009) by the American Physical Society.}
\label{f10b}
\end{figure}

A comparison of the CCE cross section (full line) with GSI data \cite{Aum99} 
is presented in \fig{f10b}. 
The disagreement already discussed for the E1 strength in \Sec{E1s} is clearly visible. 
The data do not show as large a cross section at low energies as the theory. 
It is not even clear whether the $2^+$ resonance is visible in these data. 
Nevertheless the agreement is reasonably good above 2 MeV, given that no parameter is fitted 
to this experiment in the model calculation. 
The $1^-$ contribution is calculated with two different ways of eliminating the unphysical 
bound states in the $\alpha$ + n potentials (dashed and dotted lines). 
The low-energy peak corresponds to a broad resonance in the lowest $1^-$ three-body phase shift. 
Further experimental and theoretical works are needed to explain this discrepancy.

The advantage of the relative simplicity of the CCE is that various types of angular 
differential cross sections can be calculated. 
Examples of double differential cross sections showing various energy repartitions 
between the fragments are presented in Fig.~7 of \Ref{BCD09}. 

\subsection{The CDCC approximation for three-body projectiles}
\label{cdcc3b}
The CDCC method has also been extended to three-body projectiles. 
In the first works \cite{MHO04,MEO06,RAG08}, the pseudostate discretization was adopted. 
Indeed, it avoids the difficult construction of scattering states and 
allows an accurate treatment using expansions involving Gaussians with various widths. 
Only recently was the construction of bins attempted \cite{RAG09}. 
The difficulty of the calculation restricted the first applications to elastic scattering. 

\begin{figure}
\center
\includegraphics[width=7cm]{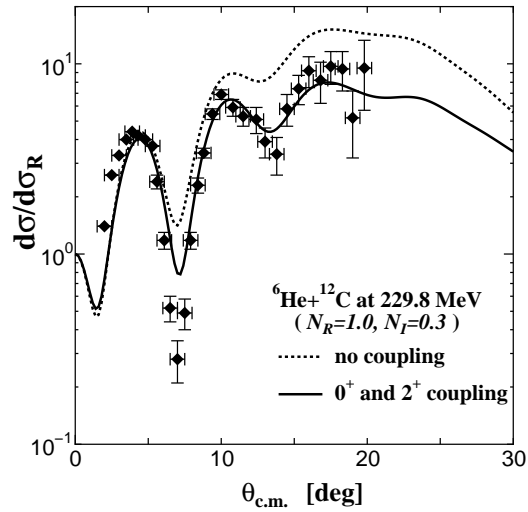}
\caption{Ratios of differential cross sections obtained with CDCC to Rutherford cross section 
for the elastic scattering of $^6$He on $^{12}$C at 229.8 MeV 
without and with coupling to breakup channels \cite{MHO04}. 
Experimental data from \Ref{LAA02}.
Reprinted figure with permission from \Ref{MHO04}.
Copyright (2004) by the American Physical Society.}
\label{f11}
\end{figure}

\begin{figure}
\center
\includegraphics[width=7cm]{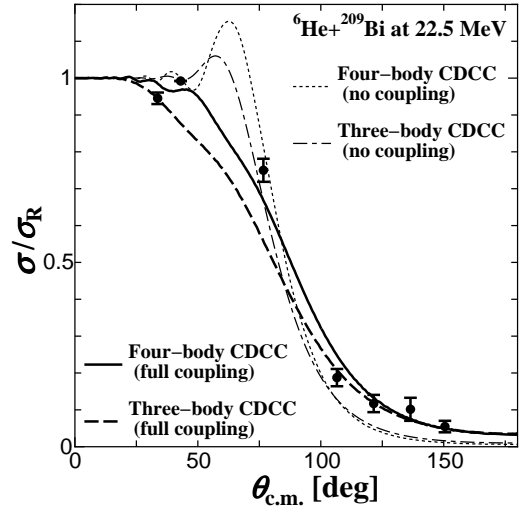}
\caption{Ratios of differential cross sections to Rutherford cross section 
for the elastic scattering of $^6$He on $^{209}$Bi at 22.5 MeV: 
comparison of three- and four-body CDCC without and with coupling to breakup channels \cite{MEO06}. 
Experimental data from Refs.~\cite{AKN00,AKC01}.
Reprinted figure with permission from \Ref{MEO06}.
Copyright (2006) by the American Physical Society.}
\label{f12}
\end{figure}

The differential cross section for elastic scattering of $^6$He on $^{12}$C 
at 229.8 MeV  is displayed in \fig{f11}. 
A single-channel calculation neglecting breakup channels (dotted line) overestimates 
the experimental data of \Ref{LAA02}. 
The shape of the data is very well reproduced by introducing $0^+$ and $2^+$ pseudochannels 
and taking account of all couplings (full line). 
In \fig{f12} is displayed a comparison between calculations of \ex{6}He elastic scattering 
on \ex{209}Bi at 22.5 MeV involving two-cluster (three-body, dashed line) 
and three-cluster (four-body, full line) descriptions of \ex{6}He. 
A significant difference appears between calculations neglecting breakup channels 
(``no coupling'') and those including it (``full coupling''). 
The agreement with experimental data \cite{AKN00,AKC01} seems better within the four-body treatment 
including the breakup channels. 


\begin{figure}
\center
\includegraphics[width=7cm]{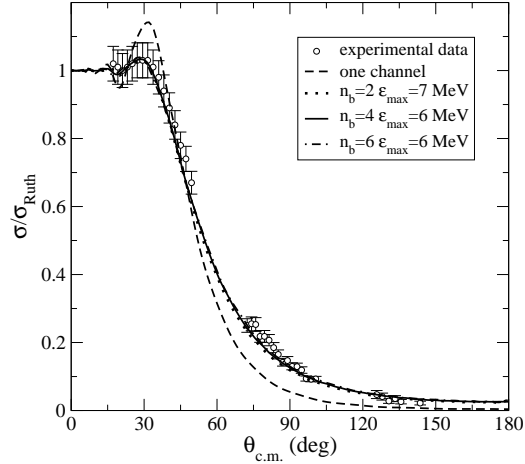}
\caption{Ratios of differential cross sections to Rutherford cross section 
for the elastic scattering of $^6$He on $^{64}$Zn at 13.6 MeV: comparison of CDCC calculations 
with various basis sizes and maximum energies $E_{\rm max}$ with a single-channel calculation \cite{RAG08}. 
Experimental data from \Ref{DFA04}.
Reprinted figure with permission from \Ref{RAG08}.
Copyright (2008) by the American Physical Society.}
\label{f13}
\end{figure}

Another type of basis functions, based on deformed oscillators, has been used to construct \ex{6}He 
pseudostates in \Ref{RAG08}. 
This technique also allowed a description of elastic scattering explicitly including 
breakup channels. 
In \fig{f13}, the elastic scattering of $^6$He on $^{64}$Zn at 13.6 MeV is compared 
with experimental data from \Ref{DFA04}. 
These results show that including partial waves up to $J = 2$ and taking coupling into 
account (full line) allow a good agreement with data (dots). 
Here also, the calculation omitting the coupling to the continuum 
(dashed line) disagrees with the experimental data. 
The same basis has recently been extended to the construction of bins \cite{RAG09}. 

While, for three-body projectiles, the effect of breakup channels has been included for some time 
in studies of elastic scattering, 
the determination of breakup cross sections is just starting. 
Some preliminary calculations have been published recently. 
Some of them are not fully converged \cite{RAG09} or involve simplifying assumptions \cite{EMO09}. 
A recent CDCC calculation \cite{MKY10} provides a good agreement with experiment \cite{Aum99} 
for $^6$He breakup on $^{12}$C. 
For $^6$He breakup on $^{208}$Pb, it does not agree well with experiment 
and is about a factor of two lower than the CCE results of \Ref{BCD09} displayed in \fig{f10b}. 
The reasons of these discrepancies are not yet understood. 
Nevertheless, the CDCC method should allow a precise treatment of three-body breakup in a near future.

\section{Perspectives}
The theory of breakup reactions offers several accurate non-relativistic approximations 
covering a broad energy range, that allow an interpretation of various experiments. 
A good accuracy is reached for some time for the breakup of two-body projectiles 
and is in view for the breakup of three-body projectiles. 
Good results can already be obtained with the simplest models of projectile structure, 
provided that the value of the projectile binding energy is correct. 
This suggests that only limited spectroscopic information can be extracted from the comparison 
of theory and experiment. 
This is partly due to the fact that a comparison of experimental data with results of calculations 
usually requires complicated convolutions. 
Nevertheless, breakup has proved to be an efficient alternative
probe to measure the separation energies of bound states of exotic nuclei.
When performed on light targets, it also provides information about the
location and width of resonances of such nuclei.
Moreover, some information about the quantum numbers of the ground state
of exotic nuclei can be assessed from breakup measurements. 
The extraction of spectroscopic factors, however, is very sensitive 
to the accuracy of the absolute normalization of experiments. 
Moreover, the sensitivity of breakup calculations to the 
description of the continuum of the projectile indicates 
that these extractions should be performed with caution. 
In addition, Coulomb breakup on heavy targets is also used to 
measure astrophysical $S$ factors. 
However the accuracy of this indirect technique is uncertain. 

Several methods can now be applied to the breakup of three-cluster projectiles 
(CDCC method, eikonal approximation, \dots). 
They will allow studying coincidence observables that are more difficult to measure 
but less sensitive to the absolute normalization of cross sections. 
They should also allow the study of correlations between the emitted fragments. 
In this respect, efforts should be made at the interface between theory and experiment 
to facilitate the transformation of the results of model calculations into 
quantities comparable with the data, taking account of the resolution and acceptances of the detection setup. 
On the theoretical side, three-cluster bound states can be obtained with good accuracy 
but the difficult treatment of the three-body continuum still requires progress. 

Attempts to improve the model description of the projectile by including excited states 
of the clusters composing the projectile 
have started with the extended CDCC.
In the future, one can expect a further improvement by using a microscopic description 
of the projectile within the microscopic cluster model \cite{WT77,Ta81,DD10},
involving effective nucleon-nucleon forces and full antisymmetrization. 
Improvements in the projectile description should first concern bound states. 
This should reduce the uncertainties appearing in non-microscopic cluster models 
because of the effective forces between the clusters in the projectile and between 
the clusters and the target. 
Using fully antisymmetrized wave functions in breakup calculations seems to be 
within reach for two-cluster projectiles. 
This approach should open the way towards ab initio descriptions of the projectile 
based on fully realistic nucleon-nucleon forces. 

All the reaction descriptions presented in this review have been 
developed within non-relativistic quantum mechanics. However, 
relativistic effects may be significant and affect the analysis of 
breakup data, even at intermediate energies of a few tens of MeV/nucleon. 
Several authors have started analyzing these effects and have 
proposed ways to take them into account in time-dependent \cite{Esb08} 
or CDCC \cite{Ber05,OB09} frameworks. Since some of the new facilities of 
radioactive-ion beams will operate at high energies (a few hundreds of 
MeV/nucleon), these effects will have to be better understood and 
incorporated in state-of-the-art reaction models.

\begin{acknowledgement}
We would like to thank P.~Descouvemont,  G.~Goldstein, V.~S.~Melezhik, F.~M.~Nunes, and Y.~Suzuki 
for fruitful collaborations concerning the development of some parts of the methods 
and applications discussed in this text. 
This text presents research results of BriX
(Belgian research initiative on exotic nuclei), the
interuniversity attraction pole programme P6/23 initiated by the Belgian-state
Federal Services for Scientific, Technical and Cultural Affairs (FSTC). 
P.~C. acknowledges the support of the F.R.S.-FNRS and of the
National Science Foundation grant PHY-0800026. 
The authors also acknowledge travel support of the 
Fonds de la Recherche Fondamentale Collective (FRFC).
\end{acknowledgement}

\newpage


\end{document}